\documentclass[prl,superscriptaddress,twocolumn, aps]{revtex4-2}
\usepackage[english]{babel}
\usepackage[T1]{fontenc}
\usepackage{amsmath,amsthm,amssymb,amsfonts,graphicx}
\usepackage{xcolor}
\usepackage{dcolumn}
\usepackage{bm}
\usepackage[colorlinks,citecolor=blue]{hyperref}
\usepackage{url}
\usepackage{braket}
\usepackage{mathrsfs}
\usepackage{dsfont}
\def \tr {\text{Tr}}
\def \T {\mathcal{T}}

\begin{document}
\title{Quantum Time Crystal Clock and its Performance}
\footnotetext{see the Appendix at the end of the manuscript, which includes Refs.~\cite{Breuer07, Wiseman93, Cilluffo2019, Manzoni17, Zhang19, Zhang18, Trivedi18, Williams, Doob, Wolpert24, prech2024}}

\author{Ludmila Viotti}
\email{lviotti@ictp.it}
\affiliation{The Abdus Salam International Center for Theoretical Physics, Strada Costiera 11, 34151 Trieste, Italy}
\author{Marcus Huber}
\affiliation{Atominstitut, Technische Universität Wien, 1020 Vienna, Austria}
\affiliation{Institute for Quantum Optics and Quantum Information - IQOQI Vienna, Austrian Academy of Sciences, Boltzmanngasse 3, 1090 Vienna, Austria}
\author{Rosario Fazio}
\affiliation{The Abdus Salam International Center for Theoretical Physics, Strada Costiera 11, 34151 Trieste, Italy}
\affiliation{Dipartimento di Fisica, Universit\`a di Napoli Federico II, Monte S. Angelo, I-80126 Napoli, Italy}
\author{Gonzalo Manzano}
\email{gonzalo.manzano@ifisc.uib-csic.es}
\affiliation{Institute for Cross-Disciplinary Physics and Complex Systems (IFISC), UIB-CSIC, E-07122 Palma de Mallorca, Spain}

\begin{abstract}
Understanding different aspects of time is at the core of many areas in theoretical physics. 
Minimal models of continuous stochastic and quantum clocks have been proposed to explore fundamental limitations on the performance of timekeeping devices. Owing to the level of complexity in the clock structure and its energy consumption, such devices show trade-offs whose characterization remains an open challenge.  
Indeed, even conceptual designs for thermodynamically efficient quantum clocks are not yet well understood. In condensed matter theory, time-crystals were found as an exciting new phase of matter, featuring oscillations in (pseudo)-equilibrium with first experimental observations appearing recently. This naturally prompts the question: \textit{can time crystals be used as quantum clocks and what is their performance from a thermodynamic perspective?} We answer this question and find that quantum time crystals are indeed genuine quantum clocks with a performance enhanced by the spontaneous breaking of time-translation symmetry.
\end{abstract}

\maketitle	

{\it Introduction}.--
In deriving ultimate performance limits for clocks from underlying quantum dynamics one is faced with the challenge that quantum physics predicts reversible dynamics and would thus not admit an arrow of time without being supplemented by measurement and statistical arguments. Indeed, in order to create an irreversible clock signal one has to generate some amount of entropy. While this thermodynamic aspect of the arrow-of-time problem is already very old~\cite{Kawai_2007,Parrondo_2009}, recent investigations have focused on its quantitative aspects by considering the operation and performance of autonomous quantum clocks (see the reviews \cite{Mitchison_2019,Antonio_2024}). Erker {\em et al.}~\cite{Erker_2017} argued that any clock will have to produce entropy to operate and that the entropy production limits the achievable accuracy of a clock. An exemplary study of a simple model therein revealed a connection to what is now known as the thermodynamic uncertainty relation (TUR)~\cite{Horowitz20}, positing that the precision of any current is bounded by the entropy production of the process (the relation between the two has also been studied experimentally by Pearson {\em et al.}~\cite{Ares21}).
For timekeeping devices one is ultimately interested in two basic properties: how often they tick in relation to other physical systems (i.e. the frequency or resolution of the clock) and the tick's accuracy with respect to a perfectly regular signal. 
While Brownian models for classical clocks~\cite{Seifert_2016}, as well as simple instances of quantum clocks~\cite{Erker_2017} seem to have their accuracy bounded by entropy production in a linear fashion, recent investigations suggest that quantum clocks may outperform this bound~\cite{Woods_2023}, even exponentially~\cite{Meier_2024} and in turn also classical TURs can be violated using quantum systems~\cite{Pta18,Agarwalla18,Potts_2021,Rignon-Bret21,Lopez23}.
There is also a fundamental trade-off between accuracy and resolution~\cite{Meier23}. In classical post-processing one can summarize events to define a new coarse grained tick, which will improve accuracy at

\vspace{-.25cm}
\noindent the cost of resolution. On the other hand, one may also improve accuracy or resolution by engineering a complex quantum clockwork. 
This presents an opportunity to evaluate the genuine quantum performance of a clock, as beating the classical accuracy-resolution trade-off is another relevant quantum signature in clocks.

It is still a major question to which extent it is possible to design clockworks that get close, or even saturate, the above fundamental bounds while providing a time reference in a macroscopic observable. In this work we will argue that employing time-crystals as clockworks may offer an answer. A Time Crystal (TC) is a recently discovered quantum phase of matter that spontaneously breaks time-translational invariance. Since the first paper by  Wilczek~\cite{wilkzek_2012} where the seminal idea was introduced, the attention on properties and regimes to observe time crystals has continued to grow.  The no-go theorem of Ref.~\cite{watanabe_2015} moved the focus towards quantum many-body systems out of equilibrium, as e.g. Floquet systems~\cite{khemani_2016a,else_2016,zhang_2017,choi_2017}. A comprehensive account of this large body of work can be found in~\cite{sacha_2017,zaletel2023colloquium,Sacha:book}. 
For the purpose of the present work, we will consider time crystalline phases that have been proposed~\cite{iemini-prl-2018,gong_2018,buca_2019,zhu_2019,Riera_Campeny_2020,hajdusek_2022,Mattes_2023,Cabot23,Carollo_2024}, and experimentally detected~\cite{Kongkhambut_2022,Carraro_Haddad_2024}, in driven-dissipative systems. Here the TC phases appear as limit cycles of certain collective variable that become stable in the  macroscopic limit, the period only depending on the coupling constants of the dynamical equation.

Besides their intrinsic interest as a new state of matter,  time crystals may also be of relevance for quantum technological applications. This is indeed a very promising avenue to explore and initial steps in this direction have been taken~\cite{montenegro2023quantumenhanced,
choi2017quantum,iemini2023,Cabot24,Gribben25,Carollo_2020}.  What we are going to argue is that the intrinsic time-translation symmetry breaking of continuous TCs make them natural candidates for high-precision clocks. 

\begin{figure}[t]
    \centering
    \includegraphics[width=0.95\linewidth]{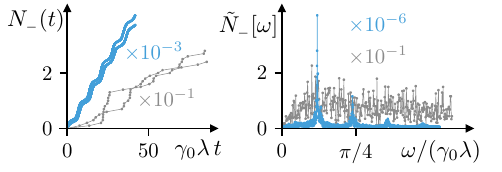}
    \includegraphics[width=0.95\linewidth]{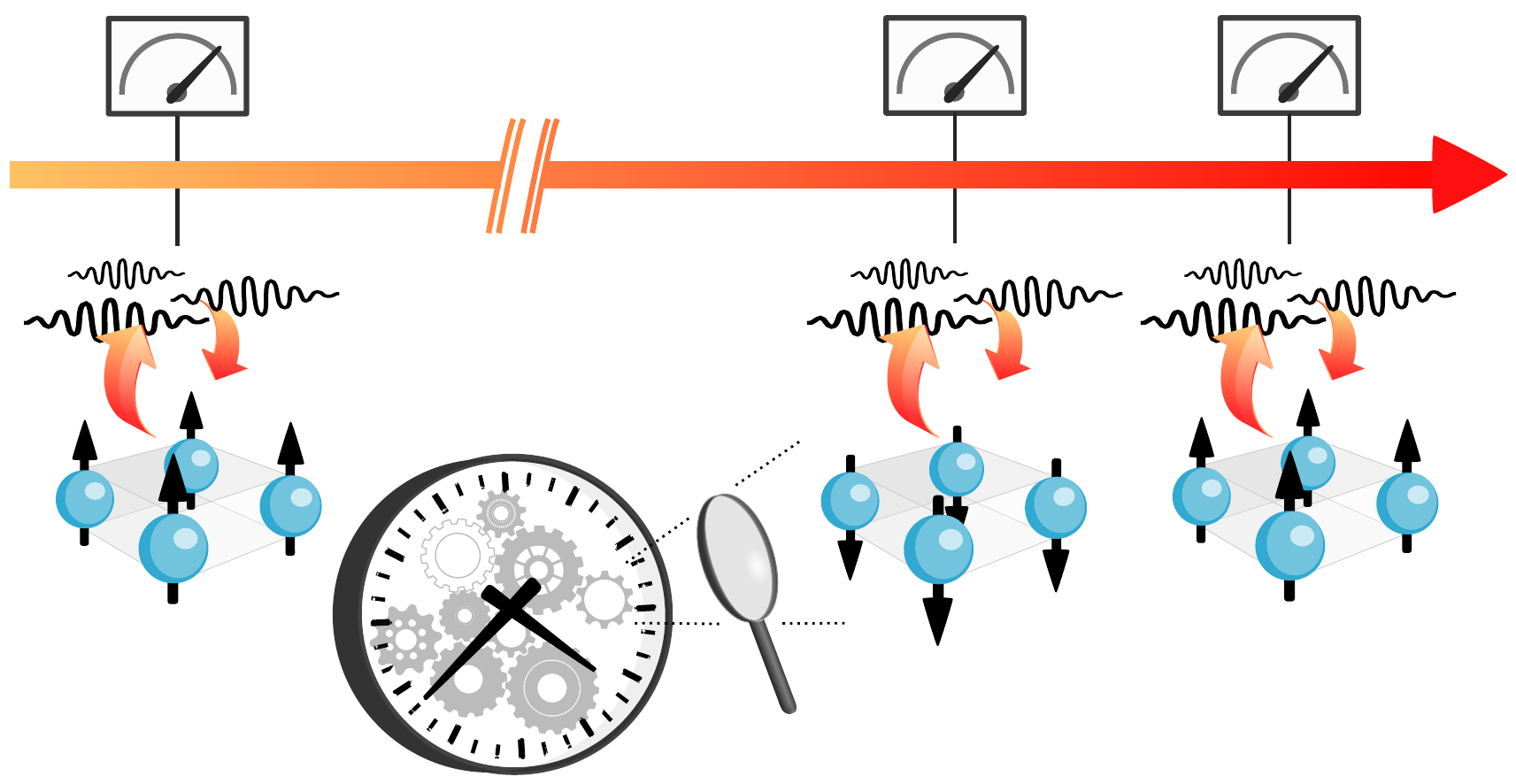}
    \caption{Illustration for the clock model composed by many identical spins collectively interacting with a non-equilibrium environment. Continuous monitoring provides a time reference via event-counting which is greatly enhanced in the time-crystal phase. In the top plot we show (left) the number (scaled by the factors indicated) of collective emissions $N_-$  accumulated along two arbitrarily chosen realizations of the dynamics, for $\lambda = 0.7$ (gray),  and $\lambda = 2$ (light blue), together with (right) the Fourier transform $\tilde{N}_-[\omega] = \int dt N_{-}${ \small$(t)$} $e^{-i\omega t}$. 
    Parameters: $S= 50$, $\gamma_0 = 10^{-3} \omega_C$ and $\beta= 2 \omega_C$.}
    \label{fig:sketch}
\end{figure}

{\it Many-body clock driven by coherence.}--We consider a clock model composed by a set of $n_s$ identical spin-1/2 particles that are collectively dissipating in a nonequilibrium bosonic environment (see the sketch in Fig.~\ref{fig:sketch}). The clock Hamiltonian reads $H_C = \sum_n \omega_C\, \sigma_n^{+} \sigma_n^{-}$, where $\sigma_n^{\pm} = (\sigma_n^{(x)} \pm i\,\sigma_n^{(y)})/2$ are operators acting on the individual spins ($n=1,2, ..., n_s$) and we take $\hbar = 1$. They interact all together with a collection of environmental bosonic modes, $H_R\!=\sum_l \omega_l\, b_l^\dagger b_l$ that are assumed to be in a thermal state at temperature $1/\beta$. 
The interaction follows the Tavis-Cummings model~\cite{Tavis67,Tavis69,Kirton19} $H_{CR}\!= \sum_l g_l\,( S_- b_l^\dagger\, e^{i \Delta_l t} + S_+ b_l\, e^{-i \Delta_l t})/\sqrt{S}$, where $S_{\pm}=S_x \pm { i}\, S_y$ and $S_i = \frac{1}{2}\sum_n \sigma_n^{(i)}$, with $i= x, y ,z$, are the collective spin operators with total spin $S=n_s/2$. The reservoir also interacts with an extra mode $H_D = \omega_C\, a^\dagger a\,$ initially prepared in a coherent state $\ket{\alpha}$, via $H_{DR} = \sum_l i\,g_l (a^\dagger\,b_l\, e^{-i \Delta_l t}-a\,b_l^\dagger\, e^{i \Delta_l t})/\sqrt{S}$. This extra mode acts as an energy source that drives the bath modes temporally out of equilibrium and induces coherence in the system, a necessary ingredient for our model to reach a time-crystalline phase in the macroscopic limit (see the Appendix~\footnotemark[1] for details).

We are interested in the clock as a monitored quantum system~\cite{Carmichael1993,Belavkin89,Dalibard1992}. The evolution of the system can be obtained within a collisional-like framework~\footnotemark[1], suitable for handling environments with quantum coherence~\cite{Manzano_2018,Rodrigues_2019,Hammam_2021,ciccarello_2022}, and unraveled using a quantum-jump trajectory approach~\cite{Manzano18,Manzano2022,Landi2024}. 
Under weak coupling and Markovian approximations, the evolution is described by the following (interaction-picture) Lindblad master equation:

\begin{equation} \label{eq:master}
  \dot{\rho}_C = \sum_{k=\pm} \frac{\gamma_k}{S} \left(L_k \rho_C L_k^{\dagger} - \frac{1}{2} \{L_k^\dagger L_k , \rho_C\} \right),
\end{equation} 
where the operators $L_\pm = S_\pm \mp i\, \alpha$ feature (collective) emissions and absorptions, further displaced by the coherent mode amplitude $\alpha\in \mathbb{R}$.  
The rates above are $\gamma_{+} = \gamma_0 \,\bar{n}$ and $\gamma_{-} = \gamma_0\, (\bar{n}+ 1)$, where $\gamma_0$ corresponds to the spontaneous emission rate and $\bar{n}=(e^{\beta \omega_C}-1)^{-1}$~\footnotemark[1]. 
As a consequence, the net effect of the non-equilibrium environment is to trigger jumps (abrupt changes in the system state) described by the displaced collective operators $L_\pm$ at rates $\gamma_\pm$ verifying a local detailed balance relation $\gamma_{\mp} = \gamma_{\pm} e^{\pm \beta \omega_C}$. This model, recently experimentally implemented~\cite{Ferioli23}, mimics the dynamics and properties of a class of dissipative time crystals~\cite{iemini-prl-2018,hajdusek_2022,Mattes_2023, Cabot23, Carollo_2024} where the period of the oscillations is not set by an external periodic source. In the macroscopic limit $S \rightarrow \infty$ (i.e, $n_s \rightarrow \infty$), the system enters in a time-crystalline phase when the parameter $\lambda \equiv \alpha / S$ exceeds a critical value $\lambda > \lambda_\mathrm{c} = 1$. It is characterized by stable oscillations of collective spin components $\langle S_i \rangle$ with frequency $\nu = (\gamma_0/2\,\pi) \sqrt{\lambda^2 - 1}$~\cite{iemini-prl-2018}. In the finite $S$ regime where the clock operates, the oscillations are damped out in the long-time limit and the dynamics governed by Eq.~\eqref{eq:master} reaches a unique nonequilibrium steady state (NESS) described by a density matrix $\pi$ satisfying $\dot{\pi} =0$.  In the NESS, input work from the extra coherent mode maintains the clock out of equilibrium at the price of dissipating heat into the bath.

\begin{figure*}[t]\centering
    \includegraphics[width=\linewidth]{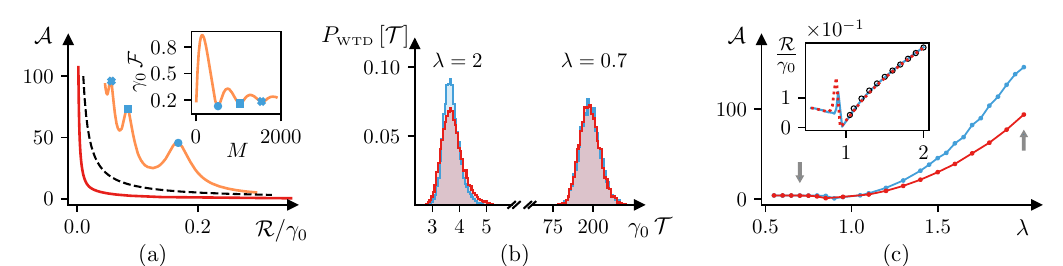}
    \vspace{-.4cm}
    \caption{ (a) Accuracy-Resolution tradeoff curves for $\lambda = 0.7$ below the critical value (red) and $\lambda = 1.5$ above the critical value (orange), in a clock with $S = 50$ ($100$ spins). The black-dashed line represents the Poissonian benchmark relation $\mathcal{A} = \mathcal{R}^{-1}$ with $\mathcal{R}$ in units of $\gamma_0$. The inset shows the Fano factor $\mathcal{F}$ (normalized with $\gamma_0$) for $\lambda = 1.5$ as a function of the threshold $M$. The optimal threshold $M = 523$ is denoted by a circular marker. The minima of $\mathcal{F}$ correspond to maxima in the tradeoff curve as signaled with same markers. (b) WTD histograms $P_{\scriptscriptstyle{\rm WTD}}[\T]$ above and below the critical point for $S = 25$ (red) and for $S = 50$ (light blue). (c) Accuracy vs. $\lambda$ (Resolution vs. $\lambda$ in the inset) at optimal threshold for $S= 25$ (red) and $S = 50$ (light blue). The gray arrows point at the $\lambda$ values for which the WTD histograms are shown on (b). The black circles in the inset show the fit to the time-crystal frequency $\mathcal{R} \sim \nu = (\gamma_0 /2\pi) \sqrt{\lambda^2 - 1}$ with coefficient $r^2= 0.999989$.
    Other parameters: $\gamma_0 = 10^{-3} \omega_C$, $\beta= 2 \omega_C$ and $N(t) = N_-(t)$.
    \label{fig:merit}}
\end{figure*}

Under continuous monitoring, the state of the system conditioned on the measurement outcomes evolves according to the stochastic Schr\"odinger equation (SSE):
\begin{equation} \label{eq:SSE}
\begin{split}
    d \ket{\psi_t} =& - \frac{dt}{2} \sum_{k= \pm} \frac{\gamma_k}{S} \left( L_k^\dagger L_k - \langle L_k^\dagger L_k \rangle \right)  \ket{\psi_t} \\ 
    &+ \sum_{k = \pm} dN_k \left( \frac{L_k \ket{\psi_t}}{\langle L_k^\dagger L_k \rangle} - \ket{\psi_t} \right),
\end{split}
\end{equation}
where we employ a direct detection scheme~\cite{Wiseman2009} for the collective jump $L_- (L_+)$, by measuring the bosons absorbed (lost) in the reservoir. Here $dN_k = \{ 0, 1 \}$ with $k= \{+,-\}$ are Poissonian stochastic increments associated with the number of jumps $N_\pm$ that are stochastically generated during the evolution, with $\langle dN_k \rangle = (\gamma_k/S)\; {\rm Tr}[L_k^\dagger L_k \pi]\,dt$, ensuring consistency with the master equation~\eqref{eq:master}. Such a description allows us to reconstruct the quantum trajectories of the spin system over its Hilbert space from the sequences of detections registered on each realization of the dynamics, and to observe the statistics of the detected jumps. 

The system retains its quantum features, as the entanglement, even in the large $S$ limit~\cite{Louren_o_2022,Mattes_2023,Passarelli_2024}.
Moreover, despite the open system evolution settles in a NESS, the monitored dynamics [Eq.~\eqref{eq:SSE}] is non-trivial and features an oscillatory behavior reminiscent of the time-crystalline phase at finite $S$. These oscillations disappear again, from their differences in phase, when taking the average over realizations.
As shown in Fig.~\ref{fig:sketch}, the impact of the time-crystalline properties of the clock become apparent in the statistics of collective jumps. 
By counting these jumps we are actually provided with a time reference that can be used to define a ticking clock. 
In particular, the clock is assumed to tick when the accumulated number of counts, $N(t) = a_- N_-(t) + a_+ N_+(t)$ for some choice of constants $a_-$ and $a_+$, reaches a predefined threshold $M$, which occurs at a stochastic time $\T$ corresponding to the minimum time verifying $N(\T) =M$. This stochastic time reference is a first passage time (FPT)~\footnote{More precisely, the stochastic time $\T \equiv \T_2 - \T_1$ is taken to be the time elapsed between two FPT corresponding to consecutive events where the threshold $M$ is reached, $\T_1$ and $\T_2$ (see the Appendix [46] for further details).}, which equals the waiting time between two consecutive ticks. The mean waiting time $\langle \T \rangle$ and its variance $\mathrm{Var}[\T] = \langle \T^2 \rangle - \langle \T \rangle^2$ can be used to characterize the clock's timekeeping performance. In particular, we define the \emph{resolution} $\mathcal{R}$, and the \emph{accuracy} $\mathcal{A}$ of the clock at FPT as:
\begin{equation}
  \mathcal{R} := \frac{1}{\langle \T \rangle} ~~~~;~~~~  \mathcal{A}:= \frac{\langle \T \rangle^2}{
   \mathrm{Var}[\T]}.
\end{equation}
The resolution measures how frequently the clock can tick and then how small the time interval it can account for is. The accuracy (or precision) of the clock is instead given by the relative dispersion of the waiting times, measuring how similar the time intervals associated to the ticks are. A third measure of dispersion that provides a compromise between accuracy and resolution is the Fano factor $\mathcal{F} = ({\mathcal{R} \mathcal{A}})^{-1}$, a noise-to-signal ratio measuring the reliability with which the waiting time random variable can be estimated. We remark that the above introduced figures of merit for the clock performance are defined at FPT, thus differing in general from the typical definitions employed in previous works on quantum clocks, based on the fluctuations of the number of counts $N(t)$~\cite{Mitchison_2019,Antonio_2024}. It should be noted that the properties of jumps discussed above, rely on the existence of a time-crystalline phase. Therefore the qualitative features apply, more generally, to the whole class of dissipative time crystals, governed by a Lindblad dynamics, that are conveniently unraveled via a stochastic dynamics as in Eq.(\ref{eq:SSE}), see e.g.~\cite{Tucker_2018,Ferioli23,Kongkhambut_2022,Dreon_2022}.

{\it Optimal thresholds.}--In any clock where ticks are defined from counting fundamental stochastic events, there is a trade-off between accuracy and resolution. If these events are independent and identically distributed grouping events always increase the accuracy, but such improvement comes at the cost of losing resolution. For Poissonian statistics with a rate $\Gamma$ (exponential waiting time distribution), we have $\mathrm{Var}[\T] = \langle \T \rangle^2 = 1/\Gamma^2$, thus $\mathcal{A} = 1$ and $\mathcal{R} = \Gamma$. Grouping $M$ of such events in a single tick leads to an increased accuracy $\mathcal{A} = M$, but resolution $\mathcal{R} = \Gamma/M$, leading to the trade-off relation $\mathcal{A} = \Gamma/ \mathcal{R}$~\cite{Meier23}. On the other hand, while the complex clockwork presented here retains the $\mathcal{R}\propto 1/M$ scaling, it shows an intricate trade-off relation between accuracy and resolution when events are grouped, leading to a nontrivial choice for thresholds $M$ optimizing the clock performance, as illustrated in Fig.~\ref{fig:merit}. In Fig.~\ref{fig:merit}a we show the resolution and accuracy tradeoff curves when varying the threshold choice $M$. We observe two qualitatively different behaviors depending on the value of the parameter $\lambda$: for $\lambda < \lambda_{\rm c}$ (red line) the tradeoff is monotonic and stays below the Poissonian tradeoff curve $\mathcal{A} = \gamma_0/ \mathcal{R}$ (dashed line). On the other hand, for $\lambda > \lambda_{\rm c}$ the tradeoff curve substantially improves (orange line), overcoming the Poissonian benchmark and acquiring structure, what points to the development of correlations in time for the counted events. Emerging peaks provide us with preferred choices of thresholds $M$ where the clock time-keeping performance is maximally enhanced. Among such choices, the peak associated with a smaller threshold - signaled with a circular marker in Fig.\ref{fig:merit}a - also minimizes the Fano factor (see inset plot). We therefore identify this as the optimal threshold, since it jointly maximizes both figures of merit while it minimizes the noise-to-signal ratio. Notably the optimal threshold scales proportionally with the system size $S$. These results suggest a quantum collective enhancement in the accuracy-resolution tradeoff not observed in other (simpler) models of quantum clocks.

The optimal threshold leads to a waiting time distribution (WTD) $P_{\scriptscriptstyle{\rm WTD}}[\T]$ that significantly differs from the case of random events, as shown in Fig.~\ref{fig:merit}b. It shrinks when the number of spins in the clock is augmented, and when increasing the value of $\lambda$, that is, when entering deeper in the time-crystal phase, it both displaces towards a smaller mean (i.e. higher resolution) and reduce its variance (i.e. higher accuracy). The situation is entirely different when considering values of $\lambda$ below the critical point,  where it becomes independent of $S$. The effect of the time-crystalline transition in the clock's performance can also be appreciated in Fig.~\ref{fig:merit}c from the prominent changes in the behavior of the accuracy and resolution of the clock when varying $\lambda$. For parameters substantially below the critical value $\lambda_{\rm c} = 1$, accuracy is small and almost constant with $\lambda$. On the other hand, we observe a superlinear increase with $\lambda$ in the time-crystalline phase, which later tends to stabilize to a linear scaling, as we will see below. Meanwhile, the resolution shows an abrupt change close to the critical point, and then increases for $\lambda > 1$ due to the long-lasting oscillatory behavior, following the scaling of the time-crystal oscillations (macroscopic-limit) frequency, $\mathcal{R} \sim \sqrt{\lambda^2 - 1}$. Moreover, in Ref.~\footnotemark[1] we show that although the resolution remains almost constant when increasing the number of spins, the accuracy scales with $S$ as $\mathcal{A} \sim \sqrt{S}$ (and hence $\mathcal{F} \sim 1/\sqrt{S}$), leading to an increasing quality of the clock when augmenting the system size~\footnote{Notice that this scaling is connected to the chosen $1/\sqrt{S}$ scaling in the interaction strength of the Tavis-Cummings model. Without scaling in the interaction strength, we expect a faster increase in the clock quality.}.

{\it Thermodynamic and kinetic performance.}-- Increasing the quality of any clock comes however with an associated thermodynamic cost, which here is evaluated by computing the entropy production at ticking times. The entropy production is a central quantity in modern formulations of the second law. It quantifies both the irreversibility and the dissipation of a fluctuating physical process in classical~\cite{Seifert2005,Kawai2007,Seifert2012} and quantum regimes~\cite{Landi2021,Manzano2022}, even for nonequilibrium reservoirs \cite{Manzano18}. 
Recent results have extended such formulation beyond fixed times to stochastic stopping times~\cite{InfStopTimes2017,QuantumMartingales19,Neri19,Gambling2021,Martingalesreview}, like the waiting times $\T$ considered here.

The entropy production per tick of the clock, can be evaluated using the so-called Martingale entropy production~\cite{QuantumMartingales19,Gambling2021}, evaluated along quantum trajectories described by Eq.~\eqref{eq:SSE}.
Assuming two consecutive ticks of the clock at stochastic times $\T_1$ and $\T_2$, the entropy production of a single tick of duration $\T = \T_2 - \T_1$ reads:
\begin{equation} \label{eq:EP}
 S_{\rm tick} (\T) =  \ln \left( \frac{\mathds{P}[\gamma_{[\T_1,\T_2]}]}{\mathds{P}[\tilde{\gamma}_{[\T_1,\T_2]}]} \right) = \Delta S_{\psi}(\T) + \beta Q(\T)
\end{equation}
where $\mathds{P}[\gamma_{[\T_1,\T_2]}]$ is the (path) probability to obtain a measurement record $\gamma_{[\T_1,\T_2]} := \{ (t_1, k_1) ... (t_M, k_M) \}$ containing a given sequence of jumps from one ticking time to the next one along the quantum trajectory, and $\tilde{\gamma}_{[\T_1,\T_2]} := \{(\T - t_M, \tilde{k}_M)...(\T - t_1, \tilde{k}_1) \}$ represents the time-reversed record, containing the inverse sequence of jumps (i.e. if $k_i ={\pm}$, $\tilde{k}_i = \mp$)~\cite{Manzano2022}. In the second equality of~\eqref{eq:EP}, $\Delta S_{\psi}(\T) = - \ln \bra{\psi_{\T_2}} \pi \ket{\psi_{\T_2}} + \ln \bra{\psi_{\T_1}} \pi \ket{\psi_{\T_1}}$ is a measure of the entropy change of the clock system along single steady-state trajectories~\footnote{We remark that the angle bracket notation in $\bra{\psi_{\T_1}} \pi \ket{\psi_{\T_1}} = \tr[ \ket{\psi_{\mathcal{T}_i}}\bra{\psi_{\mathcal{T}_i}} \pi]$ for $i=1,2$ is used to denote quantum-mechanical expectation values (and not averages over measurement records) over the NESS density matrix $\pi$.} and $Q(\T) = \omega_C [\Delta N_-(\T) - \Delta N_+(\T)]$ is the total heat dissipated in the bosonic bath, with $\Delta N_\pm(\T)= N_{\pm}(\T_2) - N_{\pm}(\T_1)$. This heat comprises both energy exchanged with the clock spins as well as work coming from the extra coherent field, $\Delta E(\mathcal{T})= W(\mathcal{T}) - Q(\mathcal{T})$, as discussed in the Appendix~\footnotemark[1].   

Although the average entropy production per tick of the clock is not a proper Kullback-Leibler divergence, the second law, $\langle S_{\rm tick}(\T) \rangle \geq 0$, is guaranteed by the waiting times fluctuation theorem for the entropy production
\begin{equation}
\langle e^{-S_{\rm tick}(\T)} \rangle = 1,    
\end{equation}
which is derived and tested in the Appendix~\footnotemark[1]. We compute the average over records $\gamma_{[\T_1,\T_2]}$ of the entropy production per tick obtained through numerical simulation of the stochastic dynamics. It shows a linear scaling with the number of spins in the clock, $\langle S_{\rm tick}(\T) \rangle \sim S$, which implies a relation of the accuracy with entropy production when varying $S$ as $\mathcal{A} \sim \sqrt{\langle S_{\rm tick}(\T) \rangle}$, as illustrated in Fig.~\ref{fig:KTUR} (see inset)~\footnote{As shown in the Appendix~[46], the dominant contribution to the entropy production per tick $\langle S_{\rm tick}(\mathcal{T}) \rangle$, comes from the heat dissipated into the thermal bath, which inherits its same scalings, $\langle Q (\mathcal{T})\rangle \sim S$}.

\begin{figure}
    \centering
    \includegraphics[width=0.90\linewidth]{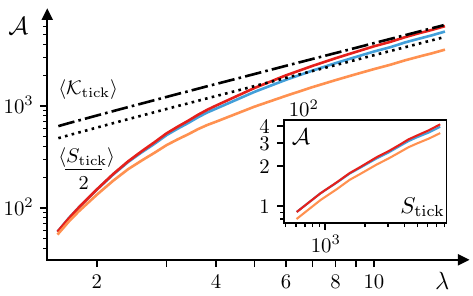}
    \caption{Accuracy curves as a function of the phase transition parameter at optimal thresholds, for $S = 50$ and different counting observables: any jump detection $N(\T) = \mathcal{K}_{\rm tick}(\T)$ (red), only emission events $N(\T) = N_-(\T)$ (light blue), and heat current $N(\T) = Q(\T)/\omega_C$ (orange). The results are compared to the TUR bound $\langle S_{\rm tick}(\T) \rangle/2$ at FPT (dotted line) and the dynamical activity $\langle \mathcal{K}_{\rm tick}(\T)\rangle$ of the KUR bound (dot-dashed line). Inset: Accuracy as a function of the average entropy production per tick when fixing $\lambda = 2$ and varying $S$. Other parameters as in Fig.~\ref{fig:sketch}.}
    \label{fig:KTUR}
    \vspace{-.4cm}
\end{figure}

When the ticking signal of the clock has the form of a current, e.g. $N(t) = Q(t)/\omega_C$, the relation of the accuracy of the clock and the average entropy production per tick are related by the FPT version of the TUR, which here reads $\mathcal{A} \leq \langle S_{\rm tick} (\T)\rangle/2$. This relation has been shown to hold for classical FPT Markov processes~\cite{Gingrich17} in the large deviation limit (i.e. for large $\T$), while a quantum counterpart is still unknown. Moreover, the accuracy of the clock for arbitrary counting observables may be related to the dynamical activity~\cite{MAES20}, by means of the Kinetic Uncertainty Relation (KUR)~\cite{Terlizzi_2019}, here reading $\mathcal{A} \leq \langle \mathcal{K}_{\rm tick} (\T)\rangle$, where the later is defined as the total number of jumps occurring during the waiting time $\T$ between two consecutive ticks, $\mathcal{K}_{\rm tick}(\T) \equiv N_-(\T) + N_+(\T)$. The KUR at FPT holds for classical Markovian processes~\cite{Garrahan17,Hiura21,Pal21} and has recently been extended to the quantum realm~\cite{VanVu22,Hasegawa22}. In Fig.~\ref{fig:KTUR} we show how both the TUR and KUR at waiting times are obeyed in our clock when the counting observable is the heat current, $N(\T) = Q(\T)/\omega_C$, while for other choices [such as counting only emissions, $N(\T) \equiv N_-(\T)$, or for the total number of jumps, $N(\T) \equiv \mathcal{K}_{\rm tick}(\T)$], the TUR is violated for large enough $\lambda$ and it is only the KUR to hold. This result implies a superiority of the time-symmetric counting observables for time keeping, which would be interesting to explore in other models as well. We also observe that, remarkably, the KUR can be saturated in the large $\lambda$ limit, when the clock operates in the far from equilibrium, deep time-crystal phase, where the oscillations become faster and more harmonic. In that regime, the scaling of the accuracy with $\lambda$ becomes linear, following the ones of the entropy production, and dynamical activity, i.e. $\mathcal{A} \sim \lambda,  \langle S_{\rm tick}(\T) \rangle \sim \lambda$, and  $\langle \mathcal{K}_{\rm tick}(\T) \rangle \sim \lambda$~\cite{Scaling}. 

{\it Final remarks.}-- 
An important aspect of the theoretical investigation into quantum clocks is the notion of autonomy \cite{Mitchison_2019,Antonio_2024}, as quantum systems driven by time-dependent classical signals, may 'hide' perfect clocks. 
There are two mathematically equivalent physical scenarios in which the TC studied in this paper emerges. Instead of coupling to a non-equilibrium environment, we could also use a classical drive on the spin system with Rabi frequency $\Omega \equiv \lambda \gamma_0$ (similarly to models in Refs.~\cite{Mattes_2023,Cabot23}). 
In our model, the coherent dynamics is instead induced by the displacement of the environmental mode acting on the thermal reservoir, which is assumed to be not directly accessible.
Despite reaching a NESS, the reminiscence of the time-crystalline phase in our model may be used as a resource to design ultra-fast and ultra-accurate quantum clocks at limited dissipation costs.

One of the crucial differences of the proposed quantum clock based on TCs to the other candidates studied in literature is the fact that the output signal used to define ticks gets more and more macroscopic as $S$ increases. While limiting the output of quantum clocks to single excitations (e.g. single photons) is beneficial in reaching ultimate performance limits of the clockwork itself, the actual dissipation of any device will have to include the measurement of the signal as well, which practically overshadows the clockwork cost by orders of magnitude~\cite{Ares_2025}. 

It is worth stressing that the performance of the TC clock studied here is not only robust against classical external noise, but the coherent behavior of the spins can be indeed used to improve the resolution and accuracy of a noisy driving signal (see Appendix~\footnotemark[1] for details). Moreover, it would be interesting to compare the thermodynamics of the present TC clock model driven by a non-equilibrium environment, with other TC models implementing similar dynamics using different resources~\cite{Carollo_2024,Carollo_2020}. Finally, it might be interesting to explore possible connections to the existing literature on time operator, see e.g.~\cite{Mu_oz_de_Nova_2025} and references therein.

\begin{acknowledgments}
We thank Albert Cabot and Roberta Zambrini for interesting discussions and comments. We acknowledge the generosity of ICTP and the CoQuSy project (No. PID2022-140506NB-C21 and C22) funded by MCIU/AEI/10.13039/501100011033. G.M. acknowledges financial support from the ``Ramón y Cajal'' program (No. RYC2021-031121-I), the QuantERA QNet project CoQuaDis (No. PCI2024-153446), and the Mar\'ia de Maeztu grant (No. CEX2021-001164-M) funded by MCIU/AEI/10.13039/501100011033 and European Union NextGenerationEU/PRTR.
We also acknowledge financial support of PNRR MUR
project PE0000023- NQSTI (L.V. and R.F.), by the European Union
ERC - RAVE, 101053159 (R.F.). MH acknowledges funding by the European Union (Quantum Flagship project ASPECTS, Grant Agreement No.\ 101080167) and the European Research Council (Consolidator grant ‘Cocoquest’ 101043705). Views and opinions expressed are however those of the author(s) only and do not necessarily reflect those of the European Union or the European Research Council. Neither the European Union nor the granting authority can be held responsible for them
\end{acknowledgments}

\vspace{.3cm}

{\textbf{Note added:} During the final stages of writing of this work, an independent work by V. Singh, E. Kwon, and G. J. Milburn appeared on the arXiv~\cite{singh2025} that studies related issues on a driven-dissipative model of cooperative resonance fluorescence at zero temperature, further supporting the relevance and increasing interest in proposing fast and accurate quantum clock models enhanced by collective effects.\nocite{data}

\vspace{-.5cm}

\bibliography{refs}

\begin{thebibliography}{100}
\bibitem [{Flu()}]{Fluctuations}%
  \BibitemOpen
  \href@noop {} {}\bibinfo {note} {Notice that the fluctuations $\delta \lambda(t)$ are conceptually important, but the tool serving the algortithm steps is the normal probability distribution according to which the $\lambda$ values are randomly generated.}\BibitemShut {Stop}%
\end{thebibliography}
\widetext
\vspace{40pt}
\begin{center}
    \textbf{\large Appendix to: ``Quantum Time-Crystal Clock and its Performance''}
\vspace{1.5cm}
\end{center}  

\twocolumngrid
\makeatletter
\setcounter{equation}{0}
\setcounter{figure}{0}
\setcounter{table}{0}
\setcounter{page}{1}
\setcounter{section}{0}
\makeatletter

\renewcommand{\thesection}{A-\Roman{section}}
\renewcommand{\thefigure}{A\arabic{figure}}
\renewcommand{\theequation}{A\arabic{equation}}
\renewcommand{\thetable}{A\arabic{table}}

\section{A-I.\hspace{.2cm}Derivation of the master equation and monitoring scheme from a non-equilibrium driven reservoir}

\subsection{Nonequilibrium environment model and dynamics}

In order to derive the dynamical evolution of the many-body system model we assume that the spins weakly interact with a nonequilibrium environment consistent of a thermal reservoir which is further coupled to an extra bosonic mode in a coherent state. The reservoir is modeled by a collection of bosonic modes with Hamiltonian $H_R = \sum_l \omega_l b_l^\dagger b_l$ ($\hbar = 1)$, where $[b_l, b_{l^\prime}^\dagger] = 
\delta_{l, l^\prime} \mathds{1}$, which are in a thermal state at inverse temperature $\beta= 1/(k_B T)$, namely, 
\begin{equation}
\rho_R = \frac{e^{-\beta H_R}}{Z_R}
\end{equation}
The extra mode is resonant with the spin natural frequency, with Hamiltonian $H_a = \omega_C\, a^\dagger a$, with $[a, a^\dagger] = \mathds{1}$. It is assumed to be in a coherent (displaced) state $\ket{\alpha}$.

The interaction of the reservoir with the spins is collective and consist in a generalized version of the Tavis-Cummings model~~\cite{Tavis67,Tavis69}. We write it as a sum over the reservoir modes 
\begin{equation}
H_{\rm int} = \sum_l H_{\rm int}^{(l)} 
\end{equation}
which are given, in the interaction picture, by
\begin{equation} \label{eq:Hint2}
\begin{split}
 H_{\mathrm{int}}^{(l)}(t) &= \frac{g_l}{\sqrt{S}} \left(S_{-} b_l^{\dagger} e^{i \Delta_l t} + S_{+} b_l e^{-i \Delta_l t} \right) \\ 
 &+ \frac{g_l}{\sqrt{S}} i \left(a b_l^\dagger e^{i \Delta_l t} -a^\dagger b_l e^{-i \Delta_l t}\right),
\end{split}
\end{equation}
where the first term corresponds to the original Tavis-Cummings model~~\cite{Tavis67,Tavis69}, with $S_\pm = S_x \pm i S_y$ collective spin operators, and the second term correspond to the interaction of the extra coherent mode with the reservoir modes. Above $\Delta_l = \omega_l - \omega_C$ is the detuning between the $l$th mode frequency $\omega_l$ and the spin system natural frequency $\omega_C$, and the parameters $g_l$ control the coupling strength of the spins (and the extra mode) to each mode $l$ in the reservoir. Here above, we included the scaling factor $1/\sqrt{S}$ in the coupling strength (with $S = n_s/2$ the total Spin) which ensures a well-behaved macroscopic limit~\cite{Kirton19}. Moreover, we assumed for simplicity same bath coupling constants $g_l$ for both the spins and the extra mode. However, it is important to mention that this assumption is not crucial for the following derivations.

The above Hamiltonian can be conveniently rewritten as:
\begin{equation} \label{eq:Hint}
 H_{\mathrm{int}}^{(l)}(t) = \frac{g_l}{\sqrt{S}} \left(L_{-} b_l^{\dagger} e^{i \Delta_l t} + L_{+} b_l e^{-i \Delta_l t} \right),
\end{equation}
in terms of the following hybrid (and collective) operators:
\begin{equation} \label{eq:jumps}
L_{-} = S_- + i a, ~~~~~~ L_{+} = S_+ - i a^\dagger
\end{equation}
verifying $L_{-} = L_+^\dagger$. Notice that these operators act simultaneously on both the spin system through the collective operators $S_{\pm}$, and on the extra (driving) environmental mode. This means that the loss (gain) of a collective excitation in the spin system is correlated with the loss (gain) of a boson in the mode $a$. 

The environmental modes are assumed to interact at random times with the spin system and the extra driving mode in a repeated manner according to the interaction Hamiltonian in Eq.~\eqref{eq:Hint}, once at a time, during a short interval of time $\tau_{\rm int}$. Such interactions are assumed to occur following Poisson statistics at a fixed rate $r_l$, which can depend on the mode. When one interaction of a mode $l$ is verified, this leads to the following unitary evolution acting on the compound mode-spins system
\begin{align} \label{eq:Uint}
 U (t + \tau_{\rm int}, t) = \exp \left( -i \int_{t}^{t + \tau_{\rm int}} H_\mathrm{int}^{(l)}(s)\,ds \right). 
\end{align}
Under the weak coupling assumption, $g_l \tau_{\rm int} \ll 1$ for all $l$, the evolution of the compound density matrix can be expanded up to second order in the coupling using the Dyson series:
\begin{align} \label{eq:Dyson}
  \rho
  (t + \tau_{\rm i}) &\simeq  \rho(t) - i \int_{t}^{t + \tau_{\rm int}} dt_1 [H_{\rm int}^{(l)}(t_1), \rho(t)] \\
   & - \int_{t}^{t + \tau_{\rm int}} dt_2 \int_{t}^{t_2} 
   \nonumber
   dt_1 [H_{\rm int}^{(l)}(t_2),[H_{\rm int}^{(l)}(t_1), \rho(t)]].
\end{align}
The first order commutator above reads:

\begin{equation}
 [H_{\rm int}^{(l)}(t_1), \rho(t)] = \frac{g_l}{\sqrt{S}}\left( [L_{-} b_l^\dagger, \rho(t)]e^{i \Delta_l t_1} + {\rm h.c.} \right)
\end{equation}
while the second order one can also be computed as:

\begin{align}
[H_{\rm int}^{(l)}(t_2), &[H_{\rm int}^{(l)}(t_1), \rho(t)]] = 
\\ \nonumber\frac{g_l^2}{S}\Big( &[L_{-} b_l^\dagger,[L_{-} b_l^\dagger,\rho(t)]]e^{i \Delta_l (t_1 + t_2)} 
\\ \nonumber+ &[L_{+} b_l,[L_{-} b_l^\dagger,\rho(t)]]e^{i \Delta_l (t_1 - t_2)} +  {\rm h.c.} \Big).
\end{align}
The reduced state of the spin system after the interaction can be obtained by partial tracing Eq.~\eqref{eq:Dyson} over all the environmental degrees of freedom. We label by $\rho_C^{(l)}(t + \tau_{\rm int}) = \tr_E[\rho(t + \tau_{\rm int})]$ the reduced state of the clock system after interaction with reservoir mode $l$, with $\tr_E$ the trace over the full environment, which is composed of the thermal reservoir and the coherent mode. Furthermore, we assume that in every interaction with the modes, the clock system always interacts with a `fresh' mode prepared in the same initial state (uncorrelated with the clock), that is, $\rho(t) = \rho_C(t) \otimes \rho_E$, with $\rho_E = \rho_R \otimes \ket{\alpha}\bra{\alpha}$. In practice, these fresh modes in the bath could be either new modes with which the spin system did not interacted yet, or the same modes of previous interactions, as long as they have thermalized again and their correlations with the spin system have decayed.

\subsection{Derivation of the master equation}

We first focus on the derivation of a quantum master equation describing the average evolution of the clock system when it is not being monitored (or, equivalently, when the information from the monitoring is not taken into account). For that purpose we consider a coarse-graining derivative for the spin system reduced evolution. During some (small) interval of time $d t \ll r_l^{-1}$ (but $d t \gg \tau_{\rm int}$), for which at most one interaction is verified, the probability that an environmental mode $l$ (with frequency $\omega_l$) interacts with the clock is $p_l = r_l\, dt$. The evolution of the system (in interaction picture) during the interval $dt$ can then be written as:

\begin{equation} \label{eq:evolution}
 \rho_C(t + dt) = \sum_{l=1}^\infty p_l\, \rho_C^{(l)}(t + \tau_{\rm int}) + (1 - \sum_{l=1}^\infty p_l)\,\rho_C(t),
\end{equation}
where the first term considers the possible interaction of the clock with the environmental modes at different frequencies and the second term accounts for the case in which no interaction takes place, leading to no evolution (free evolution) in the interaction (Schr\"odinger picture). We obtain:
\begin{align}
 \dot{\rho}_{\rm c}(t) &= \lim_{dt \rightarrow 0} \frac{\rho_C(t + dt) - \rho_C(t)}{dt} 
 \\\nonumber
 &= \sum_{l= 1}^\infty r_l \left( \rho_C^{(l)}(t + \tau_{\rm int}) - \rho_C(t) \right).
\end{align}
Performing the time integrals and the partial trace over Eq.~\eqref{eq:Dyson}, and using that the reservoir expectation values verify $\tr_R[b_l \rho_R] = \tr_R[b_l^\dagger \rho_R] = 0$, and $\tr_R[b_l^2 \rho_R] = \tr_R[b_l^{\dagger 2} \rho_R] = 0$, we obtain, after some elementary algebra, the following form for the master equation:
\begin{equation} \label{eq:master_App}
\begin{split}
\dot{\rho}_{\rm c} &= \gamma_{-} \left(L_- \rho_C L_+ - \frac{1}{2}\{ L_+ L_- , \rho_C \} \right) \\ &+ \gamma_{+} \left(L_+ \rho_C L_- - \frac{1}{2}\{ L_- L_+ , \rho_C \} \right).
\end{split}
\end{equation}
The collective operators $L_{\mp}$, after tracing also over the driving mode $a$, read:
\begin{equation} \label{eq:jumps}
L_{-} = S_- + i \alpha, ~~~~~~ L_{+} = S_+ - i \alpha^\ast
\end{equation}
where we used $\tr_a[a \ket{\alpha}\bra{\alpha}] = \alpha$. Assuming a real displacement scaling linearly with the number of spins in the system, $\alpha = \lambda S$, we recover the jump operators appearing in Eq.(1) of the main text.

Above in Eq.~\eqref{eq:master_App} we have also identified the rates appearing in the collective emission and absorption processes:
\begin{align}\label{eq:rates}
 \gamma_{-} :=& \tau_{\rm int}^2 \sum_{l= 1}^\infty r_l \frac{g_l^2}{S}   {\rm sinc}^2\left( \frac{\Delta_l \tau_{\rm int}}{2} \right) \tr_E[b_l b_l^\dagger \rho_E], \nonumber \\
  \gamma_{+} :=&  \tau_{\rm int}^2 \sum_{l= 1}^\infty r_l \frac{g_l^2}{S}   {\rm sinc}^2\left( \frac{\Delta_l \tau_{\rm int}}{2} \right) \tr_E[b_l^\dagger b_l \rho_E],
\end{align}
with ${\rm sinc}(x)= \sin(x)/x$. 
Moreover, we have neglected a Lamb-shift-like Hamiltonian term leading to a small frequency shift (as it is standard in master equation derivations~\cite{Breuer07}). 

The above rates can be computed by taking the continuous limit and using the spectral density of the reservoir $J(\omega):= \sum_l r_l \frac{g_l^2}{S} \delta(\omega - \omega_l)$, which allow us to transform the sums over the reservoir modes in integrals. For example, the emission rate $\gamma_{-}$ gives:
\begin{align} \label{eq:g-}
 \gamma_{-} &= \tau_{\rm int}^2 \int_0^\infty d\omega J(\omega) {\rm sinc}^2\left(\frac{\Delta ~\tau_{\rm int}}{2}\right) [\bar{n}(\omega) +1 ] \nonumber \\
 &\simeq \tau_{\rm int}^2 J(\omega_C) (\bar{n} + 1) = \frac{\gamma_0}{S} (\bar{n} + 1 ), 
\end{align}
where the second line follows from the fact that the square of the sinc function decays quickly with the detuning $\Delta$ and hence plays the role of a delta function inside the integral, filtering the resonant frequencies in the environment. Here we used $\bar{n}(\omega_l) = \tr_R[b_l^\dagger b_l \rho_E] = (e^{\beta \omega_l} -1)^{-1}$ for the average number of bosons in the thermal reservoir and denoted it simply by $\bar{n} \equiv \bar{n}(\omega_C)$ when evaluated at the resonant frequency. In the last equality of \eqref{eq:g-} we also identified the spontaneous emission rate in the model as $\gamma_0 = r g^2 \tau_{\rm int}^2$, with $r$ and $g$ respectively the rate and coupling strength of the resonant mode ($l$ such that $\omega_l = \omega_C$) in the thermal reservoir.

Analogously, for the emission rate accompanying the last term in Eq.~\eqref{eq:master_App}, we obtain:
\begin{align} \label{eq:g+}
 \gamma_{+}  &\simeq \tau_{\rm int}^2 J(\omega_C) \bar{n} = \frac{\gamma_0}{S} \bar{n}.
\end{align}
Therefore, the detailed balance condition reads:
\begin{equation}
\frac{\gamma_-}{\gamma_+} = \frac{\bar{n} +1}{\bar{n} } = e^{\beta \omega_C}. 
\end{equation}
With the above identifications, we recover from Eq.~\eqref{eq:master_App} the master equation~\eqref{eq:master} in the main text.

In the following we show how the Stochastic Schr\"odinger equation~\eqref{eq:SSE} in the main text can be also obtained from the microscopic model presented above by considering a continuous monitoring of the reservoir modes. This monitoring is implemented as measurements of the number of bosons in the reservoir prior and after interaction with the system and the extra mode. We also remark that, while we started our derivation from a continuous of bosonic modes in the environment, due to the ${\rm sinc}^2(x)$ functions appearing in Eqs.~\eqref{eq:rates}, only the modes in the environment resonant with the clock spin systems $\omega_C$ play a role in the effective (reduced) dynamics of the clock. Therefore, in the following derivations we will consider only resonant modes in the reservoir for simplicity.

\subsection{Monitoring scheme}

In order to monitor the changes in the reservoir modes, we consider the evolution represented by the unitary in Eq.~\eqref{eq:Uint} under a two-point-measurement scheme where projections $\{|n\rangle\langle n|\}_{n=0}^\infty$ in the $H_R$ basis are introduced before and after interaction. The evolution in Eq.~\eqref{eq:evolution} can then be rewritten as:
\begin{widetext}
\vspace{-.45cm}
\begin{align} \label{eq:evolution_2}
  \rho_C(t + dt) &= (1 - r\, dt)\,  \rho_C(t) + r\, dt \sum_{n, m}\frac{e^{-\beta \omega_C\, n}}{Z_{R_{\rm c}}} \bra{m} U \left(\rho_C(t) \otimes \ket{n}\bra{n} \right) U^\dagger \ket{m} \nonumber \\
  &=(1 - r\, dt)\,  \rho_C(t) + r\, dt \sum_{n, m} K_{m,n}\,\rho_C(t)\, K_{m, n}^\dagger,
\end{align}
where, in the second line, we introduced the measurement (Kraus) operators $K_{m, n} \equiv \sqrt{e^{-\beta \omega_C\, n}/Z_{R_{\rm c}}} \bra{m} U \ket{n}$ reading
\begin{align} \label{eq:Krausops}
K_{m, n} &=\sqrt{\frac{e^{-\beta \omega_C\, n}}{Z_{R_{\rm c}}}} \Big( \delta_{m, n} \mathds{1} - i \phi\, (L_+ \sqrt{n}\, \delta_{m, n-1} + L_- \sqrt{n+1}\, \delta_{m, n+1}) \\
&\;\;\;\;\; - \frac{\phi^2}{2} (L_+ L_- (n+1)\, \delta_{m, n} + L_-L_+ n\, \delta_{m,n}) - \frac{\phi^2}{2}(L_+^2 \sqrt{n(n-1)}\, \delta_{m, n-2} + L_-^2 \sqrt{(n+1)(n+2)}\,\delta_{m,n+2})\Big), \nonumber
\end{align}
\end{widetext}
and $\phi = g\, \tau_{\rm int}/\sqrt{S}$. These operators represent the measurement of an initial number of bosons $n$ in the reservoir mode prior to interaction, and a final number $m$. The factor $\sqrt{{e^{-\beta \omega_C n}}/{Z_{R_{\rm c}}}}$ above comes from the probability to initially find a number of bosons $n$ in the reservoir (see Ref.~\cite{Manzano18}), as corresponds to the thermal state.  

A quantum-jump unraveling can be obtained from Eq.~\eqref{eq:evolution_2} by distinguishing different events with the help of the operators $K_{n,m}$ in Eq.\eqref{eq:Krausops}. In particular we identify a new set of measurement (Kraus) operators $\{ M_i\}$ depending on a single index $i=\{0 , + ,-\}$ associated to steps of the evolution with no changes in the environment, and steps leading to jumps triggered by the emission and absorption of bosons by the environment, respectively. 

We first consider the instances that correspond to no changes in the environment. These include either no interaction between system and mode during $dt$ and same number of bosons before and after interaction ($n=m$ above). The unnormalized state of the system conditioned on no changes in the number of bosons in the reservoir during $dt$ reads: 
\begin{widetext}
\begin{align}
\bar{\rho}_{\rm c}^{(0)}(t + dt) &= (\mathds{1} - r dt)\,\rho_C(t) + r\, dt \sum_n K_{n,n}\, \rho_C(t)\, K_{n, n}^\dagger \nonumber \\ 
&= \left(\mathds{1}- \frac{r\, dt\, \phi^2}{2} (L_+ L_- (\bar{n} +1) + L_-L_+ \bar{n}\right)  \rho_C(t)\left(\mathds{1}- \frac{r\, dt\, \phi^2}{2} (L_+ L_- (\bar{n} +1) + L_-L_+ \bar{n}\right). \nonumber
\end{align}
\end{widetext}
From the equation above we can identify a ``no-jump" measurement operator $M_0$ acting as $\bar{\rho}_{\rm c}^{(0)}(t + dt) = M_0 \rho_C(t) M_0^\dagger$ with:
\begin{equation} \label{eq:M0}
\begin{split}
  M_0 &\equiv \mathds{1} - \frac{r\, dt \phi^2}{2}\left( L_+ L_- (\bar{n} + 1) + L_- L_+ \bar{n} \right) \\ &= \mathds{1} - \frac{dt}{2 S}\left( \gamma_{-} L_+ L_- + \gamma_+ L_- L_+ \right).
\end{split}
\end{equation}
The corresponding probability that no-jump occurs between times $t$ and $t+dt$ is:
\begin{align}
 p_0(t) &= \tr[M_0^\dagger M_0\, \rho_C(t)]  \\ 
 &= 1- r\, \phi^2 dt \Big(\langle L_+ L_- \rangle_t (\bar{n} + 1) + \langle L_- L_+ \rangle_t \bar{n} \Big), \nonumber \\
  &= 1 - \frac{dt}{S} \Big( \gamma_- \langle L_+ L_- \rangle_t + \gamma_+ \langle L_- L_+ \rangle_t \Big) \nonumber
\end{align}
where we denoted $\langle A \rangle_t = \tr[A \rho_C(t)]$ for any operator $A$.

Similarly, we can consider the evolution of the clock system conditioned on the fact that a boson is lost by the reservoir:
\begin{align}
\bar{\rho}_{\rm c}^{(+)}(t + dt) &= r\, dt \sum_{n} K_{n-1, n} \rho_C(t) K_{n-1, n}^\dagger \nonumber \\ &= r\,dt\, \phi\, \bar{n}\; L_+ \rho_C(t) L_-  \nonumber \\ &\equiv M_+ \rho_C(t) M_+^\dagger, 
\end{align}
from which we define a ``jump up" operator 
\begin{equation} \label{eq:M+}
M_+ \equiv \sqrt{r dt \phi \bar{n}}\, L_+ = \sqrt{dt \frac{\gamma_+}{S}} L_+     
\end{equation}
and occurs with probability 
\begin{equation}
p_+(t) = \tr[M_+^\dagger M_+ \rho_C(t)] = dt \frac{\gamma_+}{S} \langle L_+ L_- \rangle_t.
\end{equation}

Finally, the evolution of the clock system conditioned on a boson absorption in the reservoir:
\begin{align}
\bar{\rho}_{\rm c}^{(+)}(t + dt) &= r dt \sum_{n} K_{n+1, n} \rho_C(t) K_{n+1, n}^\dagger \nonumber \\ &= r\,dt\, \phi\, \bar{n}\, L_- \rho_C(t) L_+  \\ &\equiv M_- \rho_C(t) M_-^\dagger,
\end{align}
leads to a ``jump down" operator
\begin{equation} \label{eq:M-}
M_- \equiv \sqrt{r dt \phi (\bar{n} + 1)}\, L_- = \sqrt{dt \frac{\gamma_-}{S}} L_-
\end{equation}
acting with associated probability:
\begin{equation}
p_-(t) = \tr[M_-^\dagger M_- \rho_C(t)] = dt \frac{\gamma_-}{S}  \langle L_- L_+ \rangle_t.
\end{equation}
On the other hand, two-photon processes in the environment can be neglected. These processes take place with probabilities
\begin{align}
    p_{\text{-}\text{-}}(t) &= dt \frac{\gamma_-}{S}\, \frac{g^2\tau_{\rm int}^2}{2S}\,(\bar{n}+1) \langle L_-^2 L_+^2 \rangle_t\\
    p_{\scriptscriptstyle{++}}(t) &= dt \frac{\gamma_+}{S}\, \frac{g^2\tau_{\rm int}^2}{2S}\,\bar{n} \langle L_-^2 L_+^2 \rangle_t.
\end{align}
which correspond to subleading orders of the weak coupling approximation $g\,\tau_{\rm int}\ll1$ due to the extra $g^2\tau_{\rm int}^2/(2S)$ factor.
It is straightforward to check that these are well-defined measurement (Kraus) operators:
\begin{eqnarray}
    \sum_k M_k^\dagger M_k = M_0^\dagger M_0 + M_+^\dagger M_+ + M_-^\dagger M_- = \mathds{1},
\end{eqnarray}

The normalized state of the system after event $i= \{ 0, +, - \}$ has occurred during $t$ and $t+dt$ is then:
\begin{eqnarray}
{\rho}_{\rm c}^{(k)}(t + dt) = \frac{M_k \rho_C(t) M_k^\dagger}{p_k(t)},   
\end{eqnarray}
with $\sum_k p_k(t) = 1$ for all $t$. The master equation evolution in Eq.~\eqref{eq:master_App} is recovered by averaging over the three type of events defined above:
\begin{equation}
  \dot{\rho}_{\rm c}(t) = \left(\sum_{k} M_k \rho_C(t) M_k^\dagger - \rho_C(t) \right)/dt,  
\end{equation}
from which Eq.~\eqref{eq:master_App} follows upon replacing the operators $\{M_i\}$ in Eqs.~\eqref{eq:M0}, \eqref{eq:M+} and \eqref{eq:M-}.

\subsection{Derivation of the Stochastic Schr\"odinger Equation}

The stochastic Schr\"odinger equation reported in Eq.(2) of the main text can be obtained from the above derivation following a standard (direct detection) procedure~\cite{Wiseman2009,Manzano2022} with the three operators:
\begin{align}
    M_0 &:= \mathds{1} - dt (\gamma_{-}L_+ L_- + \gamma_+ L_- L_+)/(2S),  \\[.7em]
    M_- &:= \sqrt{dt \frac{\gamma_-}{S}} L_-, ~~~~   M_+ := \sqrt{dt \frac{\gamma_+}{S}} L_+,  \nonumber 
\end{align}
proportional to the shifted collective jumps $L_\mp$ in Eq.~\eqref{eq:jumps}. We define Poissonian stochastic increments $dN_k=\{0 ,1 \}$ associated to the counting of up and down events $k= \{-, +\}$, which are stochastic random variables that verify $dN_k dN_j = \delta_{k, j} dN_k$. Their average over trajectories is $\langle dN_k \rangle = \tr[M_k^\dagger M_k \rho_C(t)] = p_k(t) \propto dt$, justifying the Poissonian character of the events. Within the finite $S$ and long-time regime of operation of the clock, $\rho_C\equiv \pi$, so the mean value of the stochastic increments reads $\langle dN_k \rangle = (\gamma_k/S)\, \tr[L_k^\dagger L_k \pi]\,dt$.

Assuming a pure state of the clock at time $t$, $\ket{\psi_t}$, the conditioned evolution of the system can be written using the stochastic increments as:
\begin{align}
d \ket{\psi_t} = \left(\!1\! -\! \sum_{k=\pm} dN_k\! \right)\!\! \left(\! \frac{M_0 \ket{\psi_t}}{\sqrt{p_0(t)}} - \ket{\psi_t}\! \right)\;\, \nonumber \\ 
+ \sum_{k= \pm} dN_k \left(\!\frac{M_k \ket{\psi_t}}{\sqrt{p_k(t)}}\! -\! \ket{\psi_t}\!  \right).
\end{align}
Replacing the expressions for the operators $M_0$, $M_+$ and $M_-$ in Eqs.~\eqref{eq:M0}, \eqref{eq:M+}, and \eqref{eq:M-} respectively, we obtain:
\begin{align}
d \ket{\psi_t} &= \Big( 1\! -\!\! \sum_{k=\pm} dN_k \Big) \frac{dt}{2} \sum_{k= \pm} \frac{\gamma_k}{S} \left(\langle L_k^\dagger L_k \rangle - L_k^\dagger L_k\right)  \ket{\psi_t}  \nonumber \\  
&~+ \sum_{k= \pm} dN_k \left( \frac{L_k \ket{\psi_t}}{\langle L_k^\dagger L_k \rangle} - \ket{\psi_t} \right),    
\end{align}
where the first line describes a smooth evolution of the wavefunction $\ket{\psi_t}$ (proportional to $dt$) corresponding to the case in which no changes in the reservoir number of quanta are detected (non-zero only if $dN_+ = dN_- = 0$). Differently, the second line describes abrupt changes (i.e. jumps) in the wave-function corresponding to the cases of emission into the reservoir (non-zero when $dN_- = 1$ and $dN_+=0$) and absorption from the reservoir (non-zero if $dN_- = 0$ and $dN_+ =1$). Using that $dN_k dt \sim O(dt^2)$, the equation above can be simplified to:
\begin{align}
    d \ket{\psi_t} &= \frac{dt}{2} \sum_{k= \pm} \frac{\gamma_k}{S} \left(\langle L_k^\dagger L_k \rangle - L_k^\dagger L_k\right)  \ket{\psi_t}  \nonumber \\  
    &\;\;\;+ \sum_{k= \pm} dN_k \left( \frac{L_k \ket{\psi_t}}{\langle L_k^\dagger L_k \rangle} - \ket{\psi_t} \right),
\end{align}
which corresponds to the Stochastic Schr\"odinger Equation (2) in the main text.
Within this description, the stochastic trajectories of the spin system over its Hilbert space can be completely reconstructed from the sequences of detections $\gamma_{[0,t]}\!:=\{ (t_1, k_1)...(t_{M}, k_{ M}) \}$ registered on each realization of the dynamics. Given the equivalent content of information, in what follows we will often refer to the sequence of detections as the trajectory. 

Finally, we remark that shifted operators like those involved in this scheme are typically obtained in quantum optics monitoring setups in the case of (discrete) homodyne photodetection~\cite{Wiseman93}, where the output light of a cavity is coherently mixed in a beam splitter with a coherent field before arriving to the detector. More recently it has also been proposed the use of one-dimensional waveguides to monitor quantum jumps~\cite{Cilluffo2019, Manzoni17, Zhang19, Zhang18, Trivedi18}, where an incident field driving the system mixes, within the waveguide, with the output emitted light of the system before detection. This analogy suggest that the stochastic evolution described here might be also obtained using these platforms.

\section{A-II.\hspace{.2cm}Thermodynamics of the time crystal clock}\label{sec:sup_EP}

\subsection{Entropy production split}
In order to characterize the entropy production of the clock between ticking times, we build on the split developed in Ref.~\cite{QuantumMartingales19} for the stochastic entropy production along arbitrary quantum-jump trajectories embedded in a (virtual) two-point measurement (TPM) scheme where projective  measurements on the system $\{\Pi_i \}$ are performed at the beginning and at the end of any single stochastic evolution on the eigenbasis of the steady state density matrix $\pi = \sum_i \pi_i \Pi_i$ (see also the review in Ref.~\cite{Manzano2022}). For a trajectory $\Gamma_{[0,t]} = \{i_0, \gamma_{[0,t]}, i_t \}$ including both the initial and final outcomes $\{i_0, i_t \}$ from the (virtual) projective measurements, and the sequence of detections $\gamma_{[0,t]}\!:=\{ (t_1, k_1)...(t_{M}, k_{ M}) \}$, the (total) stochastic entropy production during the trajectory can be split as:
\begin{equation}
S_{\rm tot}(\Gamma_{[0,t]}) = S_{\rm mar}(\gamma_{[0,t]}) + S_{\rm unc}(\{i_0, i_t\}),
\end{equation}
with the two contributions named ``martingale" and ``uncertainty" entropy production respectively. The first one reads:
\begin{equation} \label{eq:mar}
 S_{\rm mar}(\gamma_{[0,t]}) = \Delta S_{\psi}(t) + \beta Q(t),   
\end{equation}
with $\Delta S_{\psi}(t) = - \ln \bra{\psi_t} \pi \ket{\psi_t} + \ln \bra{\psi_0} \pi \ket{\psi_0}$ the change in Schumacher's fidelity between the stochastic wave function satisfying Eq.(2) in the main text, and the steady state density matrix $\pi$, and $Q(t) =\omega \int_0^t dt'\, [dN_-(t') - dN_+(t')]$ is the total number of quanta entering the reservoir during the interval $[0,t]$. This contribution is extensive in time due to the dissipated heat $Q(t)$ contribution, and depends only on the jump record $\gamma_{[0,t]}$ and on the initial state of the trajectory $\ket{\psi_0}$. While the TPM scheme (in the eigenbasis of $\pi$) sets the initial wavefunction to an eigenstate of $\pi$, $\ket{\psi_0} = \ket{i_0}$, the evolution afterwards drives the stochastic wave function over the Hilbert space of the system reaching a state $\ket{\psi_t}$ at time $t$ which in general will not be an eigenstate of $\pi$ anymore. That state in general differs from $|i_t\rangle$, which corresponds to the state after (virtually) performing the second projective measurement.

The uncertainty entropy production is instead defined as:
\begin{equation}
\begin{split}
  S_{\rm unc} (\{i_0, i_t\}) &= - \ln \pi_{i_t} + \ln \bra{\psi_t} \pi \ket{\psi_t}   \\ 
  &~~~\,+ \ln \pi_{i_0} - \ln \bra{\psi_0} \pi \ket{\psi_0},
\end{split}
\end{equation}
with $\pi_{i_t} = \bra{i_t} \pi \ket{i_t}$.
It is non-extensive in time and only depends on the initial and final TPM outcomes $i_0$ and $i_t$, and on the trajectory wave-function state at initial $\ket{\psi_0}$ and final times $\ket{\psi_t}$. As discussed above, in the TPM scheme $\ket{\psi_0} = \ket{i_0}$ while in general $\ket{\psi_t} \neq \ket{i_t}$, so the second line in the uncertainty entropy vanishes, while the first line generally does not. Indeed $S_{\rm unc}$ is nonzero only when the wave-function state is not an eigenstate of the stationary density operator $\pi$ at the beginning and final points of the trajectory, what occurs as a consequence of the development of quantum superpositions on $\ket{\psi_t}$ in the $\pi$ eigenbasis during the stochastic dynamics. Therefore, $S_{\rm unc}(\{i_0, i_t \})$ vanishes in the classical case of dynamics describing ``simple" jumps among the eigenstates of $\pi$.

In Ref.~\cite{QuantumMartingales19} it has been shown that, apart from the standard fluctuation theorem for the total entropy production, $\langle e^{-S_{\rm tot}} \rangle =1$, both contributions in the above split also verify separate fluctuation theorems:
\begin{equation}
    \langle e^{-S_{\rm mar}(\gamma_{[0,t]})} \rangle = 1, ~~~~ \langle e^{-S_{\rm unc}(\gamma_{[0,t]})} \rangle = 1,
\end{equation}
which, by means of Jensen's inequality, guarantee second-law-like inequalities $\langle S_{\rm mar} \rangle \geq 0$ and $\langle S_{\rm mar} \rangle \geq 0$. As a corollary of this result, it follows that $\langle S_{\rm mar} \rangle$ is a lower bound of the total entropy production $\langle S_{\rm tot} \rangle \geq \langle S_{\rm mar} \rangle$, useful, as it only depends on the jump measurement record $\gamma_{[0,t]}$. It's also worth noticing that in the steady state $\langle S_{\rm tot} \rangle = \beta \langle Q \rangle$, while in general $\langle \Delta S_{\psi} \rangle \neq 0$ and $\langle S_{\rm unc} \rangle \neq 0$. 

\subsection{Energetics from the nonequilibrium environment}

The advantage of the continuous monitoring scheme developed here, measuring the number of bosons in the thermal reservoir prior and after interaction, is that it allows to unambiguously identify the heat flux in the setup from the net energy damped into the reservoir during an interval $[t, t +dt]$ , i.e. $dQ(t) = \omega_C (\langle dN_- \rangle - \langle dN_+ \rangle )$. This heat flux, however, involves energy coming from exchanges with the clock spins, but also with the environmental coherent mode $a$.

Indeed, it is illustrative to compute the average heat current dissipated into the thermal reservoir, which reads for a generic nonequilibrium state of the clock:
\begin{align}
 \langle \dot{Q} \rangle &= \omega_C (\langle dN_- \rangle - \langle dN_+ \rangle ) \nonumber \\
 &= \omega_C \big(\frac{\gamma_-}{S} \tr[L_+ L_- \rho_C] - \frac{\gamma_+}{S} \tr[L_- L_+ \rho_C] \big)\nonumber \\ 
 &= \omega_C \big(\frac{\gamma_-}{S} \tr[S_+ S_- \rho_C] -  \frac{\gamma_+}{S} \tr[S_- S_+ \rho_C] \big) \nonumber \\ &~~~+ \omega_C \frac{\gamma_0}{S} |\alpha|^2 - 2 \omega_C \frac{\gamma_0}{S} \alpha \tr[S_y \rho_C]
 \end{align}
Here above the first term corresponds to an energy flux directly exchanged by the spin system and the reservoir through the collective jumps $S_{\mp}$, the second term is the energy damped into the reservoir by the flux of bosons at rate $\gamma_0/S$ from the environmental coherent mode with amplitude $\alpha$, and the third term arises from the interference between the two above sources.

Noticing that the energy change in the clock can be written as: 
\begin{align}
\langle \dot{E}_{\rm c} \rangle &= \tr[H_{\rm c} \dot{\rho}_{\rm c}] \\
\nonumber
&=  \omega_C \frac{\gamma_0}{S} \alpha \tr[S_y \rho_C] \\ 
\nonumber
&~~~- \omega_C \left(\frac{\gamma_-}{S} \tr[S_+ S_- \rho_C] -  \frac{\gamma_+}{S} \tr[S_- S_+ \rho_C] \right), 
\end{align}
we use energy conservation, i.e. $\langle \dot{E}_{\rm c} \rangle = \langle \dot{W} \rangle - \langle \dot{Q} \rangle$, to identify the work performed by the nonequilibrium reservoir on the clock spins as:
\begin{align}
\langle \dot{W} \rangle &= \omega_C \frac{\gamma_0}{S} |\alpha|^2 - \omega_C \frac{\gamma_0}{S} \alpha \tr[S_y \rho_C] \nonumber \\ 
&= \omega_C \gamma_0 \lambda (\lambda S - \tr[S_y \rho_C] \rangle),
\end{align}
where in the second equality we introduced the value of the coherent mode displacement $\alpha = \lambda S$. In steady-state conditions it follows that $\langle \dot{E}_{\rm c} \rangle = 0$, and hence all the input work from the nonequilibrium environment is also dissipated as heat, $\langle \dot{W} \rangle = \langle Q \rangle$.
Under this conditions, the input energy from the extra coherent mode $\ket{\alpha}$ is used to bring the bath modes temporally out of equilibrium (during their interaction with the spins), which are then relaxed again to equilibrium at (inverse) temperature $\beta$ by dissipating heat.

Similarly, we can obtain the stochastic work performed by the non-equilibrium environment along single trajectories from the energy change in the spins, $\Delta E (t) := \langle \psi_t | H_{\rm c} | \psi_t\rangle - \langle \psi_0 | H_{\rm c} | \psi_0 \rangle$, as:
\begin{eqnarray}
    W(t) = Q(t) - \Delta E(t),
\end{eqnarray}
where $Q(t)$ is the stochastic heat damped into the bath introduced above. We remark that although on average $\langle \Delta E \rangle = 0$ in the NESS, the quantity $\Delta E(t)$ is stochastic and generically non zero along single trajectories.

\subsection{Martingale entropy production}

In this work we focus on the martingale entropy production $S_{\rm mar}(\gamma_{[0,t]})$ in Eq.~\eqref{eq:mar} which is more suitable to address first passage times, as it does not require projective measurements to be evaluated. We can conveniently rewrite Eq.~\eqref{eq:mar} as the log of a ratio of (path) probabilities of forward and time-reversed trajectories:
\begin{equation} \label{eq:Nadom}
 S_{\rm mar}(\gamma_{[0,t]}) = \ln \left( \frac{\mathds{P}[\gamma_{[0,t]}]}{\mathds{P}[\tilde{\gamma}_{[0,t]}]} \right).
\end{equation}
Here the forward path probability reads:
\begin{eqnarray}
\mathds{P}[\gamma_{[0,t]}] =  \langle \psi_0 | \pi | \psi_0 \rangle \tr[\mathcal{L}^\dagger_{{\gamma}_{[0,t]}} \mathcal{L}_{\gamma_{[0,t]}} \ket{\psi_0} \bra{\psi_0}],
\end{eqnarray}
where the first term is the probability that the system is in $\ket{\psi_0}$ at the initial time, and $\mathcal{L}_{\gamma_{[0,t]}}$ is the trajectory operator generating the sequence of jumps in the record $\gamma_{[0, t]}$. Analogously the time-reversed path probability is:
\begin{eqnarray}
\mathds{P}[\tilde{\gamma}_{[0,t]}] = \langle \psi_t | \pi | \psi_t \rangle \tr[\mathcal{L}_{\tilde{\gamma}_{[0,t]}}^{\dagger} \mathcal{L}_{\tilde{\gamma}_{[0,t]}} \ket{\psi_t} \bra{\psi_t}],
\end{eqnarray}
with $\mathcal{L}_{\tilde{\gamma}_{[0,t]}} = \mathcal{L}^\dagger_{[0,t]} e^{-\beta Q(t)/2}$ whenever the jumps occurring in the trajectory verify the local detailed balance relation~\cite{Manzano2022}, as it is the case for the time crystal clock model proposed here. For simplicity, we have assumed $\pi$ is invariant under the (anti-unitary) time-reversal operator $\Theta$. Nonetheless, the same results hold for Hamiltonians or steady-state operators that are odd under time-reversal (see e.g. Refs.~\cite{Manzano18,QuantumMartingales19, Manzano2022}).

The most prominent feature of $S_{\rm mar}(\gamma_{[0,t]})$ is that, differently from the total entropy production $S_{\rm tot}(\Gamma_{[0,t]})$, it can be used to construct an exponential Martingale, namely:
\begin{equation} \label{eq:martingale}
   \langle e^{-S_{\rm mar}(\tau)} | \gamma_{[0, t]} \rangle =  e^{-S_{\rm mar}(t)},
\end{equation}
for any $t \leq \tau$, a property that was first noticed in Ref.~\cite{QuantumMartingales19}. Martingales are a special class of stochastic processes well known in modern probability theory~\cite{Williams}, that verify their expectation value, conditioned on past observations, equals the last value observed on the process. This is to say $M(t)$ is a Martingale if and only if $\langle M(\tau) | \gamma_{[0, t]} = M(t)$ for all $\tau \geq t$. As a consequence Martingales are stochastic processes without drift.

Among the many interesting properties of Martingale processes, here we focus on their applications to first passage times or, more generally, to ``stopping times". A stopping time $\T$ is generically defined as the first time at which a certain condition (the stopping condition), evaluable over the measurement record $\gamma_{[0,\T]}$, is met. Paradigmatic examples of stopping conditions are a counting observable $N(t)$ to reach a certain threshold $M$, or a stochastic process $X(t)$ reaching a subset of the state space $\mathcal{X}$. Therefore $\T$ is a random variable that can only be evaluated from the past history $\gamma_{[0,\T]}$.

The fact that we can identify $S_{\rm mar}(t)$ as an exponential Martingale allows us to apply Doob's optional stopping theorem~\cite{Doob}. The theorem states that for any Martingale process $M(t)$, its expectation value at an arbitrary stopping time equals the expectation at the initial time $\langle M(\T) \rangle = \langle M(0) \rangle$. This is verified whenever $M(\T)$ is a well defined random variable, which can be ensured either with a bounded stopping time $\T < \infty$ (e.g. introducing a limit time in the stopping condition), or if $|M(\T)| < \infty$ when $\T \rightarrow \infty$.
Applying Doob's optional stopping theorem to $M(t) = e^{-S_{\rm mar}(t)}$, we obtain:
\begin{equation} \label{eq:stoppingFT}
    \langle e^{-S_{\rm mar}(\T)} \rangle = \langle e^{-S_{\rm mar}(0)} \rangle = 1, 
\end{equation}
which leads to an extension of the integral fluctuation theorem at generic stopping times~\cite{QuantumMartingales19}. Applying Jensen's inequality to the above fluctuation relation, it follows that $ \langle S_{\rm mar} (\T) \rangle \geq 0$ corresponds to the second law inequality for stopping times.  Notice that the average entropy production at $\T$ can be written as:
\begin{align*}
\langle S_{\rm mar} (\T) \rangle &= \int d\T P[\T] \sum_{\gamma_{[0,\T]}} \mathds{P}(\gamma_{[0,\T]} | \T) S_{\rm mar}(\gamma_{[0,\T]}) \\
&= \int d\T P[\T] \sum_{\gamma_{[0,\T]}} \mathds{P}(\gamma_{[0,\T]} | \T)  \ln \left( \frac{\mathds{P}(\gamma_{[0,\T]})}{\mathds{P}(\tilde{\gamma}_{[0,\T]})} \right),
\end{align*}
which does not correspond, in general, to a Kullback-Leibler divergence due to the presence of the stopping time distribution $P[\T]$.

These results can also be extended to a sequence of ordered stopping times $\T_1 \leq \T_2 $, (see Ref.~\cite{Wolpert24} for a related example using a Supermartingale). Since Eq.~\eqref{eq:stoppingFT} is valid for any generic stopping time $\T$ (including $\T_1$ and $\T_2$):
\begin{equation}
\langle e^{-S_\mathrm{mar} (\T_1)} \rangle = \langle e^{-S_{\rm mar}(\T_2)} \rangle.    
\end{equation}
Applying Jensen's inequality and the fact that $\T_1 \leq \T_2$, we obtain:
\begin{equation} \label{eq:chain}
  \langle S_{\rm mar} (\T_2) \rangle \geq  \langle S_{\rm mar} (\T_1) \rangle \geq 0,
\end{equation}
which can be considered as another manifestation of the second law.

\subsection{Entropy production between ticks of the clock and fluctuation theorem for waiting times}
Finally, we now apply the above results for the case of the ticking TC clock. Ticks are defined (as in the main text) as a first passage time of the counting observable $N(t)$ to reach a threshold $M$, for different choices of $N(t)$ and $M$ (see also next section for details on their computation).
In particular, in a time interval $[0,\tau]$ during which the TC clock is running, a sequence of first passage times associated to the clock ticks $\T_1, \T_2, ..., \T_i, ...$ is observed, which satisfy: 
\begin{equation}
\T_1 \leq \T_2 \leq ... \leq \T_i \leq ... \leq \tau,
\end{equation}
and correspond to the stochastic times at which the counting observable $N(t)$ reaches, for the first time, the thresholds $M, 2M, ..., iM, ...$, respectively. More precisely, here $\T_i$ are taken as the minimum between the first time at which $N(\T_i) = i M$ is satisfied, and the final time $\tau$.

We are interested in the time elapsed between two consecutive ticks, e.g. between $\T_1$ and $\T_2$. Therefore, we define our relevant stopping time as the waiting time $\T \equiv \T_2 - \T_1$. We notice that, since the clock is assumed to run in stationary conditions, the statistical properties of the waiting times are equivalent for any choice of consecutive ticks e.g. $\T_3 - \T_2$ or $\T_4 - \T_3$, etc (but not for the interval $[0,\T_1]$ leading to the first tick as recently pointed in Ref.~\cite{prech2024}). Moreover, since $\T_1$ and $\T_2$ are well defined stopping times, so is $\T$. Indeed $\T$ can be associated to the counting process $N(t)$ reaching the threshold $M$ and then being reset to $0$. 

Therefore, we can define the entropy production between two ticks of the clock as:
\begin{equation}
\begin{split}\label{eq:S_tick_sup}
  &S_{\rm tick}(\T) \equiv S_{\rm mar}(\T_2) - S_{\rm mar}(\T_1) \\ &= \Delta S_\psi(\T_2) - \Delta S_\psi(\T_1) + \beta [Q(\T_2) - Q(\T_1)] \\ &= 
  \Delta S_\psi(\T) + \beta Q(\T)
  = \ln\left( \frac{\mathds{P}[\gamma_{[\T_1,\T_2]}]}{\mathds{P}[\tilde{\gamma}_{[\T_1,\T_2]}]}\right),
\end{split}
\end{equation}
where  $\Delta S_{\psi}(\T) = - \ln \bra{\psi_{\T_2}} \pi \ket{\psi_{\T_2}} + \ln \bra{\psi_{\T_1}} \pi \ket{\psi_{\T_1}}$ is the entropy change of the clock system from $\T_1$ to $\T_2$ and $Q(\T) = \omega_C [\Delta N_-(\T) - \Delta N_+(\T)]$ is the total heat dissipated in the bosonic thermal reservoir during the tick, with $\Delta N_\pm(\T)= N_{\pm}(\T_2) - N_{\pm}(\T_1)$. We have found this second, thermal, contribution to the entropy production to be the dominant one for the cases investigated in the main text.

The average entropy production per tick verifies the second law at stopping times since $\langle  S_{\rm tick}(\T) \rangle =  \langle S_{\rm mar}(\T_2) \rangle - \langle S_{\rm mar}(\T_1) \rangle \geq 0$, where the last inequality follows from relation~\eqref{eq:chain}.

Moreover, the Martingale process $M(\T) = e^{- S_{\rm tick}(\T)}$ also verifies the waiting times fluctuation theorem [Eq.~(5) in the main text]:
\begin{equation} \label{eq:FTwait}
\langle e^{- S_{\rm tick}(\T)} \rangle = 1,
\end{equation}
which by virtue of Jensen's inequality provides an additional proof of the non-negativity of the entropy production at waiting times, $\langle S_{\rm tick}(\T) \rangle \geq 0$.

A proof of the fluctuation theorem in Eq.~\eqref{eq:FTwait} can be obtained in two steps. First:
\begin{widetext}
\begin{align} \label{eq:tickstop}
\langle e^{- S_{\rm tick}(\T)} | \T_1 \rangle &= \int_{\T_1}^\tau d\T_2 \sum_{\gamma_{[\T_1, \T_2]}} P[\T_2]~ \mathds{P}(\gamma_{[\T_1,\T_2]} | \T_2 , \T_1)  e^{- S_{\rm mar}(\T_2)}e^{S_{\rm mar}(\T_1)} \nonumber \\
&= \int_{\T_1}^{\tau} d\T_2 \sum_{\gamma_{[\T_1, \T_2]}} P[\T_2]~ \mathds{P}(\gamma_{[\T_1,\T_2]} | \T_2, \T_1) \langle e^{- S_{\rm mar}(\tau)} | \gamma_{[\T_1, \T_2]}\rangle e^{S_{\rm mar}(\T_1)}  \nonumber \\ &= \langle e^{- S_{\rm mar}(\tau) +  S_{\rm mar}(\T_1)} | \T_1 \rangle = 1,
\end{align}
\end{widetext}
where we used $\T_1 \leq \T_2 \leq \tau$ (with $\tau$ arbitrary large), the martingale property of $e^{- S_{\rm mar}(\T_2)}$ in the second line, and, in the third line, $S_{\rm mar}(\tau) - S_{\rm mar}(\T_1) = \ln \left( \frac{\mathds{P}(\gamma_{[\T_1, \tau]})}{\mathds{P}(\tilde{\gamma}_{[\T_1, \tau]})}\right)$. Note that the final equality follows from the integral fluctuation theorem at fixed times (from $\T_1$ to $\tau$), i.e. 
\begin{equation}
\begin{split}
&\langle e^{- S_{\rm mar}(\tau) +  S_{\rm mar}(\T_1)} | \T_1 \rangle \\ &= \sum_{\gamma_{[\T_1 , \tau]}} \mathds{P}(\gamma_{[\T_1 , \tau]} | \T_1) e^{-S_{\rm mar}(\tau) + S_{\rm mar}(\T_1)} \\ &= \sum_{\tilde{\gamma}_{[\T_1 , \tau]}} \mathds{P}(\tilde{\gamma}_{[\T_1 , \tau]}) = 1
\end{split}
\end{equation}
since we are conditioning on $\T_1$, which hence becomes fixed. Then the final result in Eq.~\eqref{eq:FTwait} can be simply obtained by taking the average of the above equation over the initial ticking time $\T_1$, i.e. 
\begin{equation}
\begin{split}
\langle e^{- S_{\rm tick}(\T)} \rangle &= \int_0^\tau dT_1 P[\T_1] \langle e^{- S_{\rm tick}(\T)} | \T_1 \rangle \\ &= \int_0^\tau dT_1 P[\T_1] = 1. ~~~~\Box
\end{split}
\end{equation}

\section{A-III.\hspace{.2cm}Counting ticks in simulations}
\subsection{Counting observables}\label{sec:algorithm}
As in the main text, we associate a ``tick" of the clock when the accumulated number of counts $N(t) = a_-\,N_-(t) + a_+\,N_+(t)$, for some choice of $a_\pm$, reaches a (multiple of a) predefined threshold $M$.
We focus on three particular cases: (i) ticks defined counting accumulated emissions $N(t) = N_-(t)$, what implies setting $a_- = 1$ and $a_+ = 0$, (ii) ticks defined from the dynamical activity $N(t) = \mathcal{K}(t)$, obtained if $a_- = a_+ = 1$, and (iii) ticks based on the dissipated heat current $N(t) = Q(t)/\omega_C$, resulting from setting $a_\pm = \mp \,1$.

To implement the tick counting and further extract the waiting time distribution, we simulate the stochastic dynamics described by Eq. (2) in the main text according to Quantum Monte-Carlo methods. The initial (pure) state on each trajectory is an eigenstate of $\pi = \sum_n \pi_n \Pi_n$, the NESS of the unsupervised (Lindblad) dynamics ruled by Eq. (1) in the main text, and it is sampled according to its (eigenvalue) probability $\pi_n$.

Each single trajectory is then completely determined by the initial state and the measurement record $\gamma_{[0, \tau]} = \{ (t_1, k_1) ... (t_J, k_J) \}$ containing the time instants at which collective events (emission or absorption) were detected, and the type of event that took place at each detection time, up to a final time $\tau$. Out from this string, it is straightforward to construct any accumulated number of counts $N(t)$ through the formulas introduced above.
Going along $N(t)$, we can single out a set of times $\T_i$ which satisfy that $N(\T_i) = i \,M$ for $i= 1, 2, 3, ...$, meaning that the accumulated number of counts reached a multiple $i$ of the predefined threshold $M$ for the first time. This set of first passage times are the ``ticking times" of the clock i.e. the times at which the clock ticks.

For the three types of counts mentioned above, we compute the corresponding observable $N(t)$ along the trajectory obtained from the simulation, and the set of ticking times associated to a certain threshold $M$. Changing $M$ along a suitable range, several sets of ticking times are obtained. The total length $\tau$ of the trajectories is adjusted so that even the largest threshold $M$ considered is reached at least 20 times. The set of ticking times maps into a set of waiting times $\{\T = \T_{i+1} - \T_{i}\}_\gamma$ for each individual trajectory $\gamma{[0,\tau]}$. Repeating this procedure for many trajectories, we obtain a complete ensemble of waiting times $\{\T\}$ in which the $\T$ values are sampled with frequencies corresponding to the ``waiting time distribution" $P_{\scriptscriptstyle{\rm WTD}}[\T]$. 

In addition, once a set of ticking times $\{\T_i\}_\gamma$, defined from a certain countable quantity $N(t)$ and a prefixed threshold $M$, is obtained for a trajectory $\gamma_{[0, \tau]}$, we can compute the martingale entropy production for tick-times $S_{\rm mar}(\T_i)$ as introduced in the previous Sec.~\ref{sec:sup_EP}, and the dynamical activity $\mathcal{K}(\T_i) = N_-(\T_i) + N_+(\T_i)$ informing the total number of jumps recorded during the trajectory. From $S_{\rm mar}(\T_i)$, also the entropy production per tick $S_{\rm tick}(\T) = \{S_{\rm mar}(\T_{i+1}) - S_{\rm mar}(\T_{i})\}$ in Eq.~(4) of the main text can be computed, as well as the dynamical activity per tick $\mathcal{K}_{\rm tick}(\T) = \{ \mathcal{K}(\T_{i+1}) - \mathcal{K}(\T_{i})\}$. 
Finally, merging the ticks of many trajectories $\gamma_{[0,\tau]}$, we can compute expectation values $\langle X(\T)\rangle$ and variances ${\rm Var}[X(\T)]$ of any stochastic quantity $X(\T)$ over the set of waiting times $\{\T\}$. 

\subsection{Testing the entropy production fluctuation theorem for waiting times}
Here we numerically test the fluctuation theorem for the entropy production per tick, Eq.(5) in the main text. This also serves as a consistency check for our simulations, where we also verify that the entropy values collected following the above steps satisfy the stopping-time fluctuation theorem for the martingale entropy production. 

We first consider the Martingale entropy production $S_{\rm mar}(\T_1)$ for the stopping time $\T_1$ associated to the first tick in the clock with a maximum running time $\tau$, that is, the entropy production up to $\T_1 = \min(\T | N(\T) = 1,\tau)$. Notice that this ensures $\T_1$ to be bounded since $\tau$ is the bounded maximum time of the simulation runs. Averaging $e^{-S_{\rm mar}(\T)}$ over many trajectories we find complete agreement with the fluctuation theorem in Eq.~\eqref{eq:stoppingFT}, as shown in Fig.~\ref{fig:conv_th}a, where $\langle e^{-S_{\rm mar}(\T)} \rangle$ is plotted as a function of the number of trajectories used in the simulation. Moreover, although the agreement is found independently of the countable observable used for defining the ticks, showing the convergence to $1$ requires a different amount of trajectories depending on it. In particular using ticks defined from the heat current $N(t) = Q/\omega_C$ requires many more trajectories than the two other cases studied (only emissions and dynamical activity).

\begin{figure}[t]
    \centering
    \includegraphics[width=0.95\linewidth]{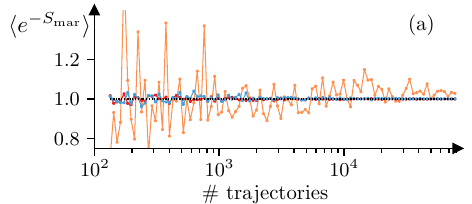}
    \includegraphics[width=0.95\linewidth]{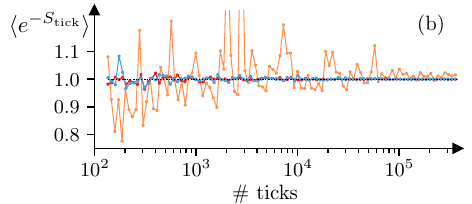}
    \caption{Convergence of the stopping-time fluctuation theorem for (a)
    $S_{\rm mar}(\T_1)$, and (b) $S_{\rm tick}(\T)$. Ticking times are defined counting collective emissions $N_-$ (light blue), dynamical activity $\mathcal{K}$ (red) and heat current $Q/\omega_C$ (orange), while setting threshold $M = 5$ counts. Parameters: The system has total spin $S=50$ ($n_s = 100$ spins) and parameter $\lambda = 2$. The environment is defined by $\gamma_0 = 10^{-3} \omega_C$ and $\beta= 0.1\, \omega_C$.} 
    \label{fig:conv_th}
\end{figure}

Turning to the entropy production per tick, $S_{\rm tick}(\T)$, a limited time for the trajectories require to use the associated (bounded) stopping times, $\T_i = \{\min (\T | N(\T) = i\,M , \tau)$. As a consequence, three computational cases arise, depending on the amount of ticks achieved during the interval $[0, \tau]$. For trajectories in which several ticks take place, e.g. $J$ ticks, we consider the tick times $\T_i = \T | N(\T) = i\,M$ with $i = 1,2, ..., J$. In trajectories where a single tick takes place, the first stopping time reads $\T_1 = \T | N(\T) = M$, while $\T_2 = \tau$. Finally, for trajectories with no ticks at all, we have $\T_1 = \T_2 = \tau$, what results in a vanishing contribution upon subtracting $S_{\rm mar}(\T_2) - S_{\rm mar}(\T_1)$.
As shown in Fig.~\ref{fig:conv_th}b, also for $S_{\rm tick}(\T)$, the fluctuation theorems converge to $1$ following the theoretical value of the theorem (Eq.~(5) in the main text). As before, while the convergence is granted for all counting observables, it is much slower when ticks are defined from $Q/\omega_C$. We further find that, for both $S_{\rm mar}$ and $S_{\rm tick}$, the amount of trajectories required to see converge to $1$ in their respective fluctuation theorems grow with $\beta$ (i.e. as temperature decreases) as well as with larger values of the threshold $M$.

\subsection{Optimal thresholds}\label{sec:bestM}
Through the procedure in Sec.~\ref{sec:algorithm} above, we obtain waiting time distributions $P_{\scriptscriptstyle{\rm WTD}}[\T]$ that depend on the threshold $M$ in an intricate manner. For instance, within the time-crystalline phase, the distribution changes back and forth between one-peaked and two-peaked distributions as $M$ is increased, as shown in Fig.~\ref{fig:Histo_th}. The alternated narrowing and broadening of the distribution induces a non-monotone variance ${\rm Var}[\T]$, so that the TC clock accuracy $\mathcal{A}$ shows a structure with maxima and minima. Meanwhile, increasing the threshold $M$ necessarily implies waiting longer times, as a greater amount of dynamical events are required for each tick. Therefore, the resolution $\mathcal{R}$ strictly decreases as a function of $M$ for any regime of the parameters.

\begin{figure}[t]
    \centering
    \includegraphics[width=0.95\linewidth]{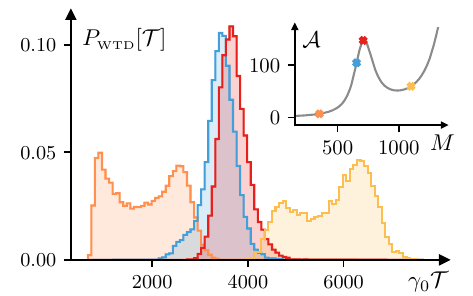}
    \caption{Waiting time distributions $P_{\rm WTD}[\T]$ obtained counting collective emissions $N_-(t)$ and fixing different threshold values $M = 355$ (orange), $M = 650$ (light blue), $M = 705$ (red), and $M = 1100$ (yellow). The inset shows the accuracy of the clock as a function of $M$. The values of $M$  associated to the shown distributions are signaled by '${\rm x}$' markers. Parameters: The system has total spin $S=50$ ($100$ spins) and parameter $\lambda = 2$. Other parameters as in Fig.~1 of the main text.} 
    \label{fig:Histo_th}
\end{figure}

In view of the intricate accuracy-to-resolution trade-off relation that emerges from this $M$-dependence, we single out an optimal threshold $M^*$ which jointly maximizes both figures of merit, while also minimizing the noise-to-ratio signal described by the Fano factor.

In the normal phase $\lambda < \lambda_{\rm c}$ the counts accumulate slowly and incoherently (see Fig.~1 in main text). The small amount of collective events results the need to wait long times until reaching the $M$ counts. Consequently, the resolution decreases fast as $\mathcal{R}\propto 1/M$ as seen in Fig.~\ref{fig:M_vs_lamb}. The linear increment observed for the accuracy $\mathcal{A}$ does not compensate the lost in resolution, leading to small optimal thresholds $M^*$ which are also independent on the system's size, as seen in the inset of Fig.~\ref{fig:M_vs_lamb}.

Over the critical value $\lambda_{\rm c} = 1$, the collective jumps drastically increase in quantity. Therefore, a fix threshold $M$ is reached much faster in the crystalline phase than in the normal phase, and the decay $\mathcal{R}\propto 1/M$ is much slower, as seen in Fig.~\ref{fig:M_vs_lamb}. The jumps also distribute inhomogeneously on time, concentrating around specific instants in a stair-like fashion (see Fig.~1 in the main text). The uneven distribution of events reveals a finite-size remnant of the time-crystalline phase, and induces the non-monotone behavior of $\mathcal{A}(M)$ described above. 
In this phase, the optimal threshold $M^*$ is proportional to the total spin $S$ and, after a short range, increases with $\lambda$ linearly as seen in the inset Fig.~\ref{fig:M_vs_lamb}. A simple linear fit indicates that
\begin{equation}\label{eq:scalingnew}
    M^*(S, \lambda) = S\,(a_M \,\lambda + b_M)
\end{equation}
obtaining $a_M = 7.31$, and $b_M = - 0.435$ and a $r^2 = 0.9999905$ coefficient, where we used values $\lambda \geq 1.3$. 

\begin{figure}[t]
    \centering
    \includegraphics[width=0.85\linewidth]{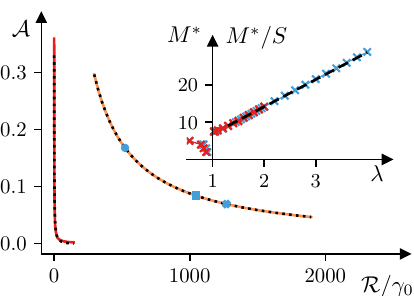}
    \caption{Resolution (normalized with $\gamma_0$) as a function of the predefined threshold $M$ for $\lambda = 0.7$ below the critical value (red) and $\lambda = 1.5$ above the critical value (orange).
    The inset shows the optimal threshold $M^*$ as a function of the parameter $\lambda$, for a system with total spin $S = 25$ (red) and $S = 50$ (light blue). For $\lambda > 1$ the threshold is shown scaled with the total spin $M^*/S$. The black-dashed line shows the linear fit performed for $\lambda \geq 1.3$. Other parameters as in Fig.~1 of the main text.\label{fig:M_vs_lamb}}
\end{figure}

As can be appreciated from the above analysis, the optimal thresholds $M^\ast$ are such that the average time between consecutive ticks roughly coincides with the time-crystal oscillation period. Physically, we interpret this result as a reminiscence of the time-crystalline phase for a finite number of spins, $n_S$. In this sense, the clock seems to work optimally by taking advantage of the natural time-crystal oscillations. Constructing on this idea, Eq.~\eqref{eq:scalingnew} tell us that the number of excitations exchanged per oscillation of the time-crystal increases linearly with $\lambda$ when entering deep in the time-crystal phase, in accordance with the fact that both the entropy production and the dynamical activity per tick also scale linearly with $\lambda$ (see Fig. 3 of the main text).

\begin{figure*}[t]
    \centering
    \includegraphics[width=\linewidth]{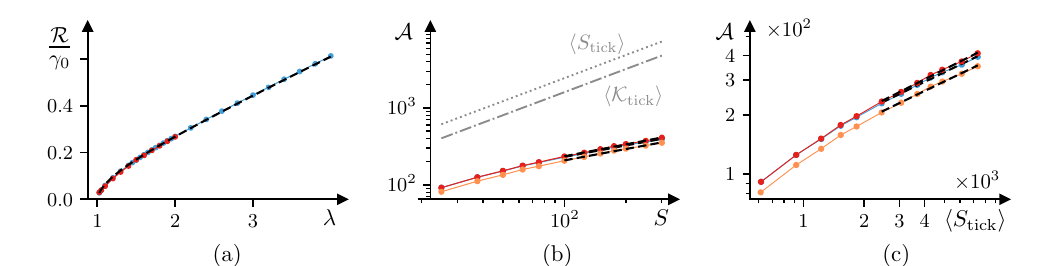}
    \caption{(a) Resolution (normalized with $\gamma_0$) as a function of $\lambda$ for systems with total spin $S = 25$ (red) and $S = 50$ (light blue), and ticks defined from collective emissions $N_-$. (b) Accuracy as a function of the total spin $S$, for a system with $\lambda =2$, and ticks defined from collective emissions $N_-$ (light blue), dynamical activity $\mathcal{K}$ (red) and heat current ${Q}/\omega_C$ (orange). The gray dotted, and dot-dashed lines show the average entropy production per tick $\langle S_{\rm tick} \rangle$ and dynamical activity per tick $\langle \mathcal{K}_{\rm tick} \rangle$, which apply for any counting observable}. (c) Accuracy as a function of $\langle S_{\rm tick} \rangle$ when parametrized with the total spin while keeping $\lambda$ fixed, same data and parameters as in (b) panel. In all the three panels, the black-dashed lines over the data points display the fits. Other parameters as in Fig.~1 of the main text.\label{fig:fits}
    \vspace{-.25cm}
\end{figure*}

\section{A-IV.\hspace{.2cm}Figures of merit scaling}
The collective effects on the TC clock performance are portrayed by the scaling of the figures of merit with $S$ and $\lambda$. Here, we provide further details on the analyses leading to the scaling forms stated in the main text. 
For each set of parameters $(S, \lambda)$, we restrict ourselves to the associated optimal performance scenario that arises upon fixing the optimal threshold $M^*$ described in Sec.~\ref{sec:bestM} above.

\subsection{Resolution}\label{sec:fit_R}
We find the resolution of the TC clock to increase with $\lambda$ in the same way as the frequency of the oscillations in the (macroscopic limit) time crystal phase, that is $\mathcal{R}(\lambda)\sim \nu(\lambda) = \gamma_0 \sqrt{\lambda^2 -1}/(2 \pi)$ and to remain independent of $S$. Fitting the $\mathcal{R}(\lambda)$ relation, as shown in Fig.~\ref{fig:fits}.a, with a function $\mathcal{R}(\lambda) = a_\mathcal{R}\,\nu(\lambda) + b_\mathcal{R}$ we obtain values $a_\mathcal{R}\sim 1$ for the scale factor and $b_\mathcal{R}\ll1$ for the displacement, sustaining the proportional form provided in the main text. The exact parameter values, together with the $r^2$ coefficient measuring the fit quality are shown in Tab.~\ref{Tab:1}, showing an excellent agreement. The curves obtained from the fit are shown in Fig.~\ref{fig:fits} as black-dashed lines over the data points.

\begin{table}[h]
    \centering
    \begin{tabular}{|l|c|c|c|}
    \hline
     &$a_\mathcal{R}$& $b_\mathcal{R}$ & $r^2$  \\\hline
     $S = 25$& $1.015$ & $-0.014$ & $0.998569$\\
     $S = 50$& $1.003$ & $-0.009$ & $0.999989$\\
     \hline
    \end{tabular}
    \caption{Fitting values for the TC clock resolution, $\mathcal{R}(\lambda) = a_\mathcal{R}\,\nu(\lambda) + b_\mathcal{R}$, and coefficient of determination $r^2$ for two different system sizes.}
    \label{Tab:1}
\end{table}

\subsection{Accuracy}
Differently from resolution, the accuracy $\mathcal{A}$ of the clock grows with the total spin $S$ (which is to say, with the system size $n_s$), as we show in Fig.~\ref{fig:fits}.b. For large enough $S \gtrsim 75$, the accuracy follows a sublinear power law $\mathcal{A}\propto S^m$, while both the dynamical activity of the clock and the entropy production per tick grow linearly with $S$. We deduce the exponent on the accuracy power-law by fitting the log-log relation with a linear function $\ln(\mathcal{A}) = m\,\ln(S) + b$, from what we get $m \simeq 1/2$. Proposing the same log-log linear functional form for $\mathcal{K}$ and $S_{\rm tick}$, we obtained $m\simeq 1$ for both quantities, confirming the linear scaling. We note that, $\mathcal{K}$ and $S_{\rm tick}$ show negligible differences when computed using different counting observables. 
Although the quantities scale in general differently with the number of counts, the changes in the corresponding optimal threshold values compensates for this difference, resulting in nearly identical values. Therefore, each is represented by a single line in Fig.~\eqref{fig:fits}. The exact values for the powers $m$ and the scale factor $b$ exacted from the fit, together with the corresponding $r^2$ coefficients are listed in Tab.~\ref{Tab:2}.

\begin{table}[h]
    \centering
    \begin{tabular}{|l|c|c|c|}
     \hline
     &$m$& $b$ & $r^2$  \\
     \hline
     $\mathcal{A}_{N_-}$& $0.494$ & $3.16$ & $0.996574$\\
     $\mathcal{A}_\mathcal{K}$& $0.513$ & $3.09$ & $0.997463$\\
     $\mathcal{A}_Q$& $0.491$ & $3.07$ & $0.996404$
     \\
     $\langle \mathcal{K}_{\rm tick} \rangle$& $0.999$ & $3.20$ & $0.999999$ \\
     $\langle S_{\rm tick} \rangle$& $0.999$ & $2.78$ & $0.999999$
     \\\hline
\end{tabular}
\caption{Fitting values for the TC clock accuracy, $\mathcal{A}(\lambda) = m\,\ln(S) + b$, and corresponding coefficients of determination $r^2$, for the three different counting observables $N(t)=\{ N_-(t), \mathcal{K}(t), Q(t)/\omega_C\}$, and for the dynamical activity $\langle \mathcal{K}_{\rm tick}\rangle$ and entropy production per tick $\langle S_{\rm tick} \rangle$.}
    \label{Tab:2}
\end{table}

The different power in the scaling of $\mathcal{A}(S)$ and $\langle S_{\rm tick} \rangle(S)$ breeds a relation $\mathcal{A}\sim\sqrt{S_{\rm tick}}$ between the accuracy of the clock and the entropy production when following curves parametrized by $S$ while keeping $\lambda$ fixed. This relation is displayed in Fig.~\ref{fig:fits}.c. For completeness, we again fit the log-log relation as $\ln(\mathcal{A}) = m \,\ln(\langle S_{\rm tick} \rangle) + b$, and verify the $m\sim 1/2$ result supporting the square root relation. The exact values obtained in the accuracy-entropy fit are given in Tab.~\ref{Tab:3}

\begin{table}[h]
    \centering
    \begin{tabular}{|l|c|c|c|}
    \hline
     &$m$& $b$ & $r^2$  \\
    \hline
     $\mathcal{A}_{N_-}$& $0.494$ & $1.58$ & $0.996725$\\
     $\mathcal{A}_\mathcal{K}$& $0.514$ & $1.45$ & $0.997502$\\
     $\mathcal{A}_Q$& $0.491$ & $1.50$ & $0.996582$\\
     \hline
\end{tabular}
\caption{Fitting values for accuracy-entropy production curves, $\ln(\mathcal{A}) = m \,\ln(\langle S_{\rm tick} \rangle) + b$, and corresponding coefficients of determination $r^2$, for the three different counting observables $N(t)=\{ N_-(t), \mathcal{K}(t), Q(t)/\omega_C\}$.}
    \label{Tab:3}
\end{table}

\section{A-V.\hspace{.2cm}Stability under Classical noise}
The dynamics described by Eq.~(1) of the main text can be derived from an alternative a microscopic model involving a classical driving instead of an additional coupling of the reservoir to a mode in a coherent state. 
Such microscopic model would be described by a clock Hamiltonian $H_C = \sum_n \omega_{\rm c}\, \sigma_n^{+} \sigma_n^{-} + V(t)$ with $V(t) = \gamma_0\,\lambda\,S_x\,\cos(\omega_C\,t)$ a classical driving of amplitude $\Omega = \gamma_0\,\lambda$, resonant with the spin natural frequency $\omega_{\rm c}$. The thermal reservoir, which is described exactly as before, constitutes in that case the only environmental degrees of freedom. The interaction between the clock and the reservoir remains the same as in the original microscopic model. Under the additional assumptions of weak-coupling to the driving $\Omega \ll \omega_{\rm c}$, and discarding the counter-rotating terms, the evolution is described by Eq.~(1) int he main text.
In this alternative scenario, it becomes possible to ask about the effect of noise in the classical driving signal. In this section, we study the stability of the (Rabi-like) oscillations induced on the system through a driving of noisy amplitude $\Omega(t) = \gamma_0\,\lambda(t)$ with $\lambda(t)$ featuring random fluctuations $\delta \lambda (t) = \lambda(t) - \langle\lambda\rangle$ around its mean value $\langle\lambda\rangle$. That is, we will define $\lambda(t)$ as an stochastic random variable normally distributed around the ideal (mean) value with variance ${\rm Var}[\lambda]$. In particular we show how the detrimental effects of the classical noise can be attenuated by the TC clock stable oscillations.
 
Amplitude fluctuations in the driving translate in Rabi oscillations with average period $\langle\T_{\rm R}\rangle = (2\pi/\gamma_0)\,\langle\lambda\rangle^{-1}$ and variance ${\rm Var}[\T_{\rm R}] =(2\pi/\gamma_0)^2\,\langle\lambda\rangle^{-4}{\rm Var}[\lambda]$. Therefore, such Rabi signal is characterized by resolution $\mathcal{R}_{\rm R}$ and accuracy $\mathcal{A}_{\rm R}$ which depend on $\lambda$ as
\begin{equation}
    \mathcal{R}_{\rm R} = \frac{\gamma_0}{2\pi}\langle\lambda\rangle\hspace{.5cm};\hspace{.5cm}\mathcal{A}_{\rm R} = \frac{\langle\lambda\rangle^2}{{\rm Var}[\lambda]}.
    \label{eq:merit_rabi}
\end{equation}
In the following we compare these quantities with the resolution and accuracy of the TC clock signal under noisy driving conditions.

To accommodate classical noise in the simulation of the stochastic dynamics, the procedure presented in Sec.~\ref{sec:algorithm}  needs to be adapted. That said, once the dynamics is simulated, the counting procedure goes exactly the same.
The main tool for implementing the dynamics is the definition of the normal probability distribution of mean $\langle \lambda \rangle$ and variance ${\rm Var}[\lambda]$ \cite{Fluctuations} that will replace the former fixed value $\lambda$. 
 For each trajectory simulation we randomly generate a $\lambda$ value according to this probability distribution and compute the NESS $\pi_\lambda$ of the unsupervised dynamics, so as to initialize the trajectory in an eigenstate of $\pi_\lambda$, which accounts for potential temporal fluctuations, while ensuring that the ensemble-averaged dynamics still recover $\pi$. In addition, the displaced jump operators become time-dependent through $\lambda$, i.e. $L_\pm (t) = S_{\pm} \mp i\,\lambda(t) S$, where $\lambda(t)$ is a sequence of values randomly sampled according to the same normal distribution. From this point on, the procedure goes exactly as in Sec.~\ref{sec:algorithm}.

\begin{figure}[h]
    \centering
    \includegraphics[width=0.95\linewidth]{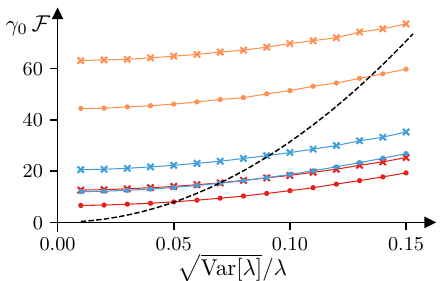}
    \caption{Fano factor $\mathcal{F}$ of the TC clock (normalized with $\gamma_0$) under noisy driving as a function of the (relative) noise amplitude $\sqrt{{\rm Var}[\lambda]}/\lambda$ for ticks defined from collective emissions $N_-$ (light blue), dynamical activity $\mathcal{K}$ (red) and heat current ${Q}/\omega_C$ (orange).
    The black-dashed line shows the Fano factor $\mathcal{F}_{\rm R}$ associated with the Rabi oscillations.
    {Parameters: The system is characterized by total spin $S = 25$ (`{\rm x}' markers) or $S = 50$ (dot markers) and mean $\langle \lambda\rangle = 2$. Other parameters are as in Fig.~1 of the main text.}\label{fig:Noise}}
    \vspace{-.25cm}
\end{figure}

We find the fluctuations on the driving signal to have the main effect of lowering the clock's accuracy without altering its form as a function of the threshold $M$. For this reason, the optimal threshold $M^*$ remains basically unchanged from the case of ideal driving, and the resolution $\mathcal{R}$ of the clock, which depends only on the mean waiting time and on the predefined threshold, stays robust to the inclusion of the classical noise.

Following the results in Sec.~\ref{sec:fit_R} and taking $\mathcal{R}\sim\nu(\lambda)$ the clock's resolution is related to that of the imperfect Rabi oscillations in Eq.~(\ref{eq:merit_rabi}) as $\mathcal{R} \sim \mathcal{R}_{\rm R}\,\sqrt{\lambda^2-1}/\lambda$. Therefore, the clock signal presents lower resolution that tends to the driving resolution as $\lambda\rightarrow\infty$. 

On the other hand, the accuracy $\mathcal{A}_{\rm R}$ of the Rabi signal can be either larger or smaller than the one of the clock, causing an inversion on the signals' performance - as quantified by the Fano factor - when increasing the relative noise amplitude $\sqrt{{\rm Var}[\lambda]}/\lambda$. More specifically, below a critical noise amplitude, we have better performance of the imperfect drive oscillations $\mathcal{F}_{\rm R}< \mathcal{F}$ while above that value, the relation inverts and the TC clock performs better than the original driving (see Fig.~\ref{fig:Noise}). The specific critical value of $\sqrt{{\rm Var}[\lambda]}/\lambda$ at which the TC clock outperforms the driving depends both on $S$ and on the type of counts used to define the ticks of the clock. 

Larger systems perform better, accelerating the TC clock out-performance. Meanwhile, clocks using the dynamical activity to count ticks show better performance, with Fano factors going below $\mathcal{F}_{\rm R}$ for smaller noise amplitudes, as compared to clocks using heat currents to count ticks, which require much bigger noise amplitudes to overcome the imperfect driving. For instance, for parameters $\langle\lambda\rangle = 2$ and $S = 50$ as considered in Fig.~\ref{fig:Noise}, we show that a noise of $5\%$ is enough for a $\mathcal{K}$-based TC clock to outperform the driving signal, which remains more stable than that of the $Q$-based TC clock for much bigger fluctuations $\sim 12\%$.

\end{document}